\newif\ifojtemplate
\ojtemplatefalse

\ifojtemplate
  \documentclass{IEEEoj}
\else
  \documentclass[10pt,journal,letterpaper]{IEEEtran}
\fi

\usepackage{url}
\usepackage[utf8]{inputenc}
\usepackage{amsmath}
\usepackage{amssymb}
\usepackage{xspace}
\usepackage{epsfig}
\usepackage{balance}
\usepackage{listings}
\usepackage[dvipsnames]{xcolor}

\AtBeginDocument{\definecolor{ojcolor}{cmyk}{0.93,0.59,0.15,0.02}}

\definecolor{codegray}{rgb}{0.25,0.25,0.25}
\definecolor{codepurple}{rgb}{0.58,0,0.82}
\lstdefinestyle{mystyle-yaml}{
  commentstyle=\color{gray},
  keywordstyle=\color{purple},
  numberstyle=\tiny\color{codegray},
  stringstyle=\color{codepurple},
  basicstyle=\color{Periwinkle}\ttfamily\scriptsize,
  rulecolor=\color{black},
  breakatwhitespace=true,         
  breaklines=true,                 
  captionpos=b,
  frame=tb,
  keepspaces=true,                 
  numbers=left,                    
  numbersep=5pt,                  
  showspaces=false,                
  showstringspaces=false,
  showtabs=false,                  
  tabsize=2,
  xleftmargin=10pt,
  aboveskip=-10pt,
  belowskip=-5pt,
}

\lstdefinelanguage{yaml}{
  alsoletter={-},
  keywords={true,false,null,y,n,-},
  sensitive=false,
  comment=[l]{\#},
  morecomment=[s]{/*}{*/},
  moredelim=[l][\color{orange}]{\&},
  moredelim=[l][\color{magenta}]{*},
  moredelim=**[il][\color{purple}{:}\color{MidnightBlue}]{:},   
  morestring=[b]',
  morestring=[b]",
}

\usepackage[acronyms,nonumberlist,nopostdot,nomain,nogroupskip,acronymlists={hidden}]{glossaries}
\newglossary[algh]{hidden}{acrh}{acnh}{Hidden Acronyms}

\usepackage{booktabs}
\usepackage{tabularx}

\usepackage{tikz}
\usepackage{pgfplots}
\pgfplotsset{compat=newest}
\pgfplotsset{plot coordinates/math parser=false}
\newlength\fheight
\newlength\fwidth
\usetikzlibrary{plotmarks,patterns,decorations.pathreplacing,backgrounds,calc,arrows,arrows.meta,spy,matrix,scopes}
\usepgfplotslibrary{patchplots,groupplots}
\usepackage{tikzscale}
\usepackage[draft]{hyperref}

\newif\ifexttikz
\exttikzfalse

\ifexttikz
	\usetikzlibrary{external}
\fi

\usepackage{multirow}
\usepackage[font=footnotesize]{subcaption}
\usepackage[font=footnotesize]{caption}

\usepackage{mathtools}
\usepackage[numbers,sort&compress]{natbib}

\usepackage{soul}

\newacronym{3gpp}{3GPP}{3rd Generation Partnership Project}
\newacronym{4g}{4G}{4th generation}
\newacronym{5g}{5G}{5th generation}
\newacronym{6g}{6G}{6th generation}
\newacronym{5gc}{5GC}{5G Core}
\newacronym{adc}{ADC}{Analog to Digital Converter}
\newacronym{aerpaw}{AERPAW}{Aerial Experimentation and Research Platform for Advanced Wireless}
\newacronym{ai}{AI}{Artificial Intelligence}
\newacronym{aimd}{AIMD}{Additive Increase Multiplicative Decrease}
\newacronym{am}{AM}{Acknowledged Mode}
\newacronym{amc}{AMC}{Adaptive Modulation and Coding}
\newacronym{amf}{AMF}{Access and Mobility Management Function}
\newacronym{aops}{AOPS}{Adaptive Order Prediction Scheduling}
\newacronym{api}{API}{Application Programming Interface}
\newacronym{apn}{APN}{Access Point Name}
\newacronym{ap}{AP}{Application Protocol}
\newacronym{aqm}{AQM}{Active Queue Management}
\newacronym{ausf}{AUSF}{Authentication Server Function}
\newacronym{avc}{AVC}{Advanced Video Coding}
\newacronym{awgn}{AGWN}{Additive White Gaussian Noise}
\newacronym{balia}{BALIA}{Balanced Link Adaptation Algorithm}
\newacronym{bbu}{BBU}{Base Band Unit}
\newacronym{bdp}{BDP}{Bandwidth-Delay Product}
\newacronym{ber}{BER}{Bit Error Rate}
\newacronym{bf}{BF}{Beamforming}
\newacronym{bler}{BLER}{Block Error Rate}
\newacronym{brr}{BRR}{Bayesian Ridge Regressor}
\newacronym{bs}{BS}{Base Station}
\newacronym{bsr}{BSR}{Buffer Status Report}
\newacronym{bss}{BSS}{Business Support System}
\newacronym{ca}{CA}{Carrier Aggregation}
\newacronym{caas}{CaaS}{Connectivity-as-a-Service}
\newacronym{cb}{CB}{Code Block}
\newacronym{cc}{CC}{Congestion Control}
\newacronym{ccid}{CCID}{Congestion Control ID}
\newacronym{cco}{CC}{Carrier Component}
\newacronym{cdd}{CDD}{Cyclic Delay Diversity}
\newacronym{cdf}{CDF}{Cumulative Distribution Function}
\newacronym{cdn}{CDN}{Content Distribution Network}
\newacronym{cir}{CIR}{Channel Impulse Response}
\newacronym{cli}{CLI}{Command-line Interface}
\newacronym{cn}{CN}{Core Network}
\newacronym{codel}{CoDel}{Controlled Delay Management}
\newacronym{comac}{COMAC}{Converged Multi-Access and Core}
\newacronym{cord}{CORD}{Central Office Re-architected as a Datacenter}
\newacronym{cornet}{CORNET}{COgnitive Radio NETwork}
\newacronym{cosmos}{COSMOS}{Cloud Enhanced Open Software Defined Mobile Wireless Testbed for City-Scale Deployment}
\newacronym{cots}{COTS}{Commercial Off-the-Shelf}
\newacronym{cp}{CP}{Control Plane}
\newacronym{cyp}{CP}{Cyclic Prefix}
\newacronym{up}{UP}{User Plane}
\newacronym{cpu}{CPU}{Central Processing Unit}
\newacronym{cqi}{CQI}{Channel Quality Information}
\newacronym{cr}{CR}{Cognitive Radio}
\newacronym{cran}{CRAN}{Cloud \gls{ran}}
\newacronym{crs}{CRS}{Cell Reference Signal}
\newacronym{csi}{CSI}{Channel State Information}
\newacronym{csirs}{CSI-RS}{Channel State Information Reference Signal}
\newacronym{cu}{CU}{Central Unit}
\newacronym{cucp}{CU-CP}{Central Unit Control Plane}
\newacronym{cuup}{CU-UP}{Central Unit User Plane}
\newacronym{d2tcp}{D$^2$TCP}{Deadline-aware Data center TCP}
\newacronym{d3}{D$^3$}{Deadline-Driven Delivery}
\newacronym{dac}{DAC}{Digital to Analog Converter}
\newacronym{dag}{DAG}{Directed Acyclic Graph}
\newacronym{das}{DAS}{Distributed Antenna System}
\newacronym{dash}{DASH}{Dynamic Adaptive Streaming over HTTP}
\newacronym{dc}{DC}{Dual Connectivity}
\newacronym{dccp}{DCCP}{Datagram Congestion Control Protocol}
\newacronym{dce}{DCE}{Direct Code Execution}
\newacronym{dci}{DCI}{Downlink Control Information}
\newacronym{dctcp}{DCTCP}{Data Center TCP}
\newacronym{dl}{DL}{Downlink}
\newacronym{dmr}{DMR}{Deadline Miss Ratio}
\newacronym{dmrs}{DMRS}{DeModulation Reference Signal}
\newacronym{drlcc}{DRL-CC}{Deep Reinforcement Learning Congestion Control}
\newacronym{drs}{DRS}{Discovery Reference Signal}
\newacronym{du}{DU}{Distributed Unit}
\newacronym{e2e}{E2E}{end-to-end}
\newacronym{earfcn}{EARFCN}{E-UTRA Absolute Radio Frequency Channel Number}
\newacronym{ecaas}{ECaaS}{Edge-Cloud-as-a-Service}
\newacronym{ecn}{ECN}{Explicit Congestion Notification}
\newacronym{edf}{EDF}{Earliest Deadline First}
\newacronym{embb}{eMBB}{Enhanced Mobile Broadband}
\newacronym{empower}{EMPOWER}{EMpowering transatlantic PlatfOrms for advanced WirEless Research}
\newacronym{enb}{eNB}{evolved Node Base}
\newacronym{endc}{EN-DC}{E-UTRAN-\gls{nr} \gls{dc}}
\newacronym{epc}{EPC}{Evolved Packet Core}
\newacronym{eps}{EPS}{Evolved Packet System}
\newacronym{es}{ES}{Edge Server}
\newacronym{etsi}{ETSI}{European Telecommunications Standards Institute}
\newacronym[firstplural=Estimated Times of Arrival (ETAs)]{eta}{ETA}{Estimated Time of Arrival}
\newacronym{eutran}{E-UTRAN}{Evolved Universal Terrestrial Access Network}
\newacronym{faas}{FaaS}{Function-as-a-Service}
\newacronym{fapi}{FAPI}{Functional Application Platform Interface}
\newacronym{fdd}{FDD}{Frequency Division Duplexing}
\newacronym{fdm}{FDM}{Frequency Division Multiplexing}
\newacronym{fdma}{FDMA}{Frequency Division Multiple Access}
\newacronym{fed4fire}{FED4FIRE+}{Federation 4 Future Internet Research and Experimentation Plus}
\newacronym{fir}{FIR}{Finite Impulse Response}
\newacronym{fit}{FIT}{Future \acrlong{iot}}
\newacronym{fpga}{FPGA}{Field Programmable Gate Array}
\newacronym{fr2}{FR2}{Frequency Range 2}
\newacronym{fs}{FS}{Fast Switching}
\newacronym{fscc}{FSCC}{Flow Sharing Congestion Control}
\newacronym{ftp}{FTP}{File Transfer Protocol}
\newacronym{fw}{FW}{Flow Window}
\newacronym{ge}{GE}{Gaussian Elimination}
\newacronym{gnb}{gNB}{Next Generation Node Base}
\newacronym{gop}{GOP}{Group of Pictures}
\newacronym{gpr}{GPR}{Gaussian Process Regressor}
\newacronym{gpu}{GPU}{Graphics Processing Unit}
\newacronym{gtp}{GTP}{GPRS Tunneling Protocol}
\newacronym{gtpc}{GTP-C}{GPRS Tunnelling Protocol Control Plane}
\newacronym{gtpu}{GTP-U}{GPRS Tunnelling Protocol User Plane}
\newacronym{gtpv2c}{GTPv2-C}{\gls{gtp} v2 - Control}
\newacronym{gw}{GW}{Gateway}
\newacronym{harq}{HARQ}{Hybrid Automatic Repeat reQuest}
\newacronym{hetnet}{HetNet}{Heterogeneous Network}
\newacronym{hh}{HH}{Hard Handover}
\newacronym{hol}{HOL}{Head-of-Line}
\newacronym{hqf}{HQF}{Highest-quality-first}
\newacronym{hss}{HSS}{Home Subscription Server}
\newacronym{http}{HTTP}{HyperText Transfer Protocol}
\newacronym{ia}{IA}{Initial Access}
\newacronym{iab}{IAB}{Integrated Access and Backhaul}
\newacronym{ic}{IC}{Incident Command}
\newacronym{ietf}{IETF}{Internet Engineering Task Force}
\newacronym{imsi}{IMSI}{International Mobile Subscriber Identity}
\newacronym{imt}{IMT}{International Mobile Telecommunication}
\newacronym{iot}{IoT}{Internet of Things}
\newacronym{ip}{IP}{Internet Protocol}
\newacronym{itu}{ITU}{International Telecommunication Union}
\newacronym{kpi}{KPI}{Key Performance Indicator}
\newacronym{kpm}{KPM}{Key Performance Measurement}
\newacronym{kvm}{KVM}{Kernel-based Virtual Machine}
\newacronym{los}{LOS}{Line-of-Sight}
\newacronym{lsm}{LSM}{Link-to-System Mapping}
\newacronym{lstm}{LSTM}{Long Short Term Memory}
\newacronym{lte}{LTE}{Long Term Evolution}
\newacronym{lxc}{LXC}{Linux Container}
\newacronym{m2m}{M2M}{Machine to Machine}
\newacronym{mac}{MAC}{Medium Access Control}
\newacronym{manet}{MANET}{Mobile Ad Hoc Network}
\newacronym{mano}{MANO}{Management and Orchestration}
\newacronym{mc}{MC}{Multi-Connectivity}
\newacronym{mcc}{MCC}{Mobile Cloud Computing}
\newacronym{mchem}{MCHEM}{Massive Channel Emulator}
\newacronym{mcs}{MCS}{Modulation and Coding Scheme}
\newacronym{mec}{MEC}{Multi-access Edge Computing}
\newacronym{mec2}{MEC}{Mobile Edge Cloud}
\newacronym{mfc}{MFC}{Mobile Fog Computing}
\newacronym{mgen}{MGEN}{Multi-Generator}
\newacronym{mi}{MI}{Mutual Information}
\newacronym{mib}{MIB}{Master Information Block}
\newacronym{miesm}{MIESM}{Mutual Information Based Effective SINR}
\newacronym{mimo}{MIMO}{Multiple Input, Multiple Output}
\newacronym{ml}{ML}{Machine Learning}
\newacronym{mlr}{MLR}{Maximum-local-rate}
\newacronym[plural=\gls{mme}s,firstplural=Mobility Management Entities (MMEs)]{mme}{MME}{Mobility Management Entity}
\newacronym{mmtc}{mMTC}{Massive Machine-Type Communications}
\newacronym{mmwave}{mmWave}{millimeter wave}
\newacronym{mpdccp}{MP-DCCP}{Multipath Datagram Congestion Control Protocol}
\newacronym{mptcp}{MPTCP}{Multipath TCP}
\newacronym{mr}{MR}{Maximum Rate}
\newacronym{mrdc}{MR-DC}{Multi \gls{rat} \gls{dc}}
\newacronym{mse}{MSE}{Mean Square Error}
\newacronym{mss}{MSS}{Maximum Segment Size}
\newacronym{mt}{MT}{Mobile Termination}
\newacronym{mtd}{MTD}{Machine-Type Device}
\newacronym{mtu}{MTU}{Maximum Transmission Unit}
\newacronym{mumimo}{MU-MIMO}{Multi-user \gls{mimo}}
\newacronym{mvno}{MVNO}{Mobile Virtual Network Operator}
\newacronym{nalu}{NALU}{Network Abstraction Layer Unit}
\newacronym{nas}{NAS}{Network Attached Storage}
\newacronym{nat}{NAT}{Network Address Translation}
\newacronym{nbiot}{NB-IoT}{Narrow Band IoT}
\newacronym{nfv}{NFV}{Network Function Virtualization}
\newacronym{nfvi}{NFVI}{Network Function Virtualization Infrastructure}
\newacronym{ni}{NI}{Network Interfaces}
\newacronym{nic}{NIC}{Network Interface Card}
\newacronym{nlos}{NLOS}{Non-Line-of-Sight}
\newacronym{now}{NOW}{Non Overlapping Window}
\newacronym{nsm}{NSM}{Network Service Mesh}
\newacronym[type=hidden]{nr}{NR}{New Radio}
\newacronym{nextg}{NextG}{Next Generation}
\newacronym{nrf}{NRF}{Network Repository Function}
\newacronym{nsa}{NSA}{Non Stand Alone}
\newacronym{nse}{NSE}{Network Slicing Engine}
\newacronym{nssf}{NSSF}{Network Slice Selection Function}
\newacronym{o2i}{O2I}{Outdoor to Indoor}
\newacronym{oai}{OAI}{OpenAirInterface}
\newacronym{oaicn}{OAI-CN}{\gls{oai} \acrlong{cn}}
\newacronym{oairan}{OAI-RAN}{\acrlong{oai} \acrlong{ran}}
\newacronym{oam}{OAM}{Operations, Administration and Maintenance}
\newacronym{ofdm}{OFDM}{Orthogonal Frequency Division Multiplexing}
\newacronym{olia}{OLIA}{Opportunistic Linked Increase Algorithm}
\newacronym{omec}{OMEC}{Open Mobile Evolved Core}
\newacronym{onap}{ONAP}{Open Network Automation Platform}
\newacronym{onf}{ONF}{Open Networking Foundation}
\newacronym{onos}{ONOS}{Open Networking Operating System}
\newacronym{oom}{OOM}{\gls{onap} Operations Manager}
\newacronym{opnfv}{OPNFV}{Open Platform for \gls{nfv}}
\newacronym[type=hidden]{oran}{O-RAN}{Open \gls{ran}}
\newacronym{orbit}{ORBIT}{Open-Access Research Testbed for Next-Generation Wireless Networks}
\newacronym{os}{OS}{Operating System}
\newacronym{osm2}{OSM}{Open Street Map}
\newacronym{oss}{OSS}{Operations Support System}
\newacronym{pa}{PA}{Position-aware}
\newacronym{pase}{PASE}{Prioritization, Arbitration, and Self-adjusting Endpoints}
\newacronym{pawr}{PAWR}{Platforms for Advanced Wireless Research}
\newacronym{pbch}{PBCH}{Physical Broadcast Channel}
\newacronym{pcef}{PCEF}{Policy and Charging Enforcement Function}
\newacronym{pcfich}{PCFICH}{Physical Control Format Indicator Channel}
\newacronym{pcrf}{PCRF}{Policy and Charging Rules Function}
\newacronym{pdcch}{PDCCH}{Physical Downlink Control Channel}
\newacronym{pdcp}{PDCP}{Packet Data Convergence Protocol}
\newacronym{pdsch}{PDSCH}{Physical Downlink Shared Channel}
\newacronym{pdu}{PDU}{Packet Data Unit}
\newacronym{pf}{PF}{Proportional Fair}
\newacronym{pgw}{PGW}{Packet Gateway}
\newacronym{phich}{PHICH}{Physical Hybrid ARQ Indicator Channel}
\newacronym{phy}{PHY}{Physical}
\newacronym{pl}{PL}{Path Loss}
\newacronym{pmch}{PMCH}{Physical Multicast Channel}
\newacronym{pmi}{PMI}{Precoding Matrix Indicators}
\newacronym{powder}{POWDER}{Platform for Open Wireless Data-driven Experimental Research}
\newacronym{ppo}{PPO}{Proximal Policy Optimization}
\newacronym{ppp}{PPP}{Poisson Point Process}
\newacronym{prach}{PRACH}{Physical Random Access Channel}
\newacronym{prb}{PRB}{Physical Resource Block}
\newacronym{psnr}{PSNR}{Peak Signal to Noise Ratio}
\newacronym{pss}{PSS}{Primary Synchronization Signal}
\newacronym{pucch}{PUCCH}{Physical Uplink Control Channel}
\newacronym{pusch}{PUSCH}{Physical Uplink Shared Channel}
\newacronym{qam}{QAM}{Quadrature Amplitude Modulation}
\newacronym{qci}{QCI}{\gls{qos} Class Identifier}
\newacronym{qoe}{QoE}{Quality of Experience}
\newacronym{qos}{QoS}{Quality of Service}
\newacronym{quic}{QUIC}{Quick UDP Internet Connections}
\newacronym{rach}{RACH}{Random Access Channel}
\newacronym{ran}{RAN}{Radio Access Network}
\newacronym[firstplural=Radio Access Technologies (RATs)]{rat}{RAT}{Radio Access Technology}
\newacronym{rbg}{RBG}{Resource Block Group}
\newacronym{rcn}{RCN}{Research Coordination Network}
\newacronym{rc}{RC}{RAN Control}
\newacronym{rec}{REC}{Radio Edge Cloud}
\newacronym{red}{RED}{Random Early Detection}
\newacronym{renew}{RENEW}{Reconfigurable Eco-system for Next-generation End-to-end Wireless}
\newacronym{rf}{RF}{Radio Frequency}
\newacronym{rfc}{RFC}{Request for Comments}
\newacronym{rfr}{RFR}{Random Forest Regressor}
\newacronym{ric}{RIC}{\gls{ran} Intelligent Controller}
\newacronym{rlc}{RLC}{Radio Link Control}
\newacronym{rlf}{RLF}{Radio Link Failure}
\newacronym{rlnc}{RLNC}{Random Linear Network Coding}
\newacronym{rmr}{RMR}{RIC Message Router}
\newacronym{rmse}{RMSE}{Root Mean Squared Error}
\newacronym{rnis}{RNIS}{Radio Network Information Service}
\newacronym{rr}{RR}{Round Robin}
\newacronym{rrc}{RRC}{Radio Resource Control}
\newacronym{rrm}{RRM}{Radio Resource Management}
\newacronym{rru}{RRU}{Remote Radio Unit}
\newacronym{rs}{RS}{Remote Server}
\newacronym{rsrp}{RSRP}{Reference Signal Received Power}
\newacronym{rsrq}{RSRQ}{Reference Signal Received Quality}
\newacronym{rss}{RSS}{Received Signal Strength}
\newacronym{rssi}{RSSI}{Received Signal Strength Indicator}
\newacronym{rt}{RT}{Real-time}
\newacronym{rtt}{RTT}{Round Trip Time}
\newacronym{ru}{RU}{Radio Unit}
\newacronym{rw}{RW}{Receive Window}
\newacronym{rx}{RX}{Receiver}
\newacronym{s1ap}{S1AP}{S1 Application Protocol}
\newacronym{sa}{SA}{standalone}
\newacronym{sack}{SACK}{Selective Acknowledgment}
\newacronym{sap}{SAP}{Service Access Point}
\newacronym{sc2}{SC2}{Spectrum Collaboration Challenge}
\newacronym{scef}{SCEF}{Service Capability Exposure Function}
\newacronym{sch}{SCH}{Secondary Cell Handover}
\newacronym{scoot}{SCOOT}{Split Cycle Offset Optimization Technique}
\newacronym{sctp}{SCTP}{Stream Control Transmission Protocol}
\newacronym{sdap}{SDAP}{Service Data Adaptation Protocol}
\newacronym{sdk}{SDK}{Software Development Kit}
\newacronym{sdm}{SDM}{Space Division Multiplexing}
\newacronym{sdma}{SDMA}{Spatial Division Multiple Access}
\newacronym{sdn}{SDN}{Software-defined Networking}
\newacronym{sdr}{SDR}{Software-defined Radio}
\newacronym{seba}{SEBA}{SDN-Enabled Broadband Access}
\newacronym{sgsn}{SGSN}{Serving GPRS Support Node}
\newacronym{sgw}{SGW}{Service Gateway}
\newacronym{si}{SI}{Study Item}
\newacronym{sib}{SIB}{Secondary Information Block}
\newacronym{sinr}{SINR}{Signal to Interference plus Noise Ratio}
\newacronym{sip}{SIP}{Session Initiation Protocol}
\newacronym{siso}{SISO}{Single Input, Single Output}
\newacronym{sla}{SLA}{Service Level Agreement}
\newacronym{sm}{SM}{Service Model}
\newacronym{smf}{SMF}{Session Management Function}
\newacronym{smo}{SMO}{Service Management and Orchestration}
\newacronym{sms}{SMS}{Short Message Service}
\newacronym{smsgmsc}{SMS-GMSC}{\gls{sms}-Gateway}
\newacronym{snr}{SNR}{Signal-to-Noise-Ratio}
\newacronym{son}{SON}{Self-Organizing Network}
\newacronym{sptcp}{SPTCP}{Single Path TCP}
\newacronym{srb}{SRB}{Service Radio Bearer}
\newacronym{srn}{SRN}{Standard Radio Node}
\newacronym{srs}{SRS}{Sounding Reference Signal}
\newacronym{ss}{SS}{Synchronization Signal}
\newacronym{sss}{SSS}{Secondary Synchronization Signal}
\newacronym{st}{ST}{Spanning Tree}
\newacronym{svc}{SVC}{Scalable Video Coding}
\newacronym{tb}{TB}{Transport Block}
\newacronym{tcp}{TCP}{Transmission Control Protocol}
\newacronym{tdd}{TDD}{Time Division Duplexing}
\newacronym{tdl}{TDL}{Tapped Delay Line}
\newacronym{tdm}{TDM}{Time Division Multiplexing}
\newacronym{tdma}{TDMA}{Time Division Multiple Access}
\newacronym{tfl}{TfL}{Transport for London}
\newacronym{tfrc}{TFRC}{TCP-Friendly Rate Control}
\newacronym{tft}{TFT}{Traffic Flow Template}
\newacronym{tgen}{TGEN}{Traffic Generator}
\newacronym{tip}{TIP}{Telecom Infra Project}
\newacronym{tm}{TM}{Transparent Mode}
\newacronym{to}{TO}{Telco Operator}
\newacronym{tr}{TR}{Technical Report}
\newacronym{trp}{TRP}{Transmitter Receiver Pair}
\newacronym{ts}{TS}{Technical Specification}
\newacronym{tti}{TTI}{Transmission Time Interval}
\newacronym{ttt}{TTT}{Time-to-Trigger}
\newacronym{tx}{TX}{Transmitter}
\newacronym{uas}{UAS}{Unmanned Aerial System}
\newacronym{uav}{UAV}{Unmanned Aerial Vehicle}
\newacronym{udm}{UDM}{Unified Data Management}
\newacronym{udp}{UDP}{User Datagram Protocol}
\newacronym{udr}{UDR}{Unified Data Repository}
\newacronym{ue}{UE}{User Equipment}
\newacronym{uhd}{UHD}{\gls{usrp} Hardware Driver}
\newacronym{ul}{UL}{Uplink}
\newacronym{um}{UM}{Unacknowledged Mode}
\newacronym{uml}{UML}{Unified Modeling Language}
\newacronym{upa}{UPA}{Uniform Planar Array}
\newacronym{upf}{UPF}{User Plane Function}
\newacronym{urllc}{URLLC}{Ultra Reliable and Low Latency Communications}
\newacronym{usa}{U.S.}{United States}
\newacronym{usim}{USIM}{Universal Subscriber Identity Module}
\newacronym{usrp}{USRP}{Universal Software Radio Peripheral}
\newacronym{utc}{UTC}{Urban Traffic Control}
\newacronym{vim}{VIM}{Virtualization Infrastructure Manager}
\newacronym{vm}{VM}{Virtual Machine}
\newacronym{vnf}{VNF}{Virtual Network Function}
\newacronym{volte}{VoLTE}{Voice over \gls{lte}}
\newacronym{voltha}{VOLTHA}{Virtual OLT HArdware Abstraction}
\newacronym{vr}{VR}{Virtual Reality}
\newacronym{vran}{vRAN}{Virtualized \gls{ran}}
\newacronym{vss}{VSS}{Video Streaming Server}
\newacronym{wbf}{WBF}{Wired Bias Function}
\newacronym{wf}{WF}{Waterfilling}
\newacronym{wg}{WG}{Working Group}
\newacronym{wi}{WI}{Wireless InSite}
\newacronym{wlan}{WLAN}{Wireless Local Area Network}
\newacronym{osm}{OSM}{Open Source \gls{nfv} Management and Orchestration}
\newacronym{pnf}{PNF}{Physical Network Function}
\newacronym{drl}{DRL}{Deep Reinforcement Learning}
\newacronym{mtc}{MTC}{Machine-type Communications}
\newacronym{mns}{MnS}{Management Services}
\newacronym{ves}{VES}{\gls{vnf} Event Stream}
\newacronym{ei}{EI}{Enrichment Information}
\newacronym{fh}{FH}{Fronthaul}
\newacronym{fft}{FFT}{Fast Fourier Transform}
\newacronym{laa}{LAA}{Licensed-Assisted Access}
\newacronym{plfs}{PLFS}{Physical Layer Frequency Signals}
\newacronym{ptp}{PTP}{Precision Time Protocol}
\newacronym{cbrs}{CBRS}{Citizen Broadband Radio Service}
\newacronym{otic}{OTIC}{Open Testing and Integration Center}
\newacronym{sba}{SBA}{Service-Based Architecture}
\newacronym{cif}{CI}{cyberinfrastructure}
\newacronym{sonic}{SONiC}{Software for Open Networking in the Cloud}
\newacronym{ocp}{OCP}{Open Compute Project}
\newacronym{snmp}{SNMP}{Simple Network Management Protocol}
\newacronym{raid}{RAID}{redundant array of independent disks}
\newacronym{nfs}{NFS}{Network File Storage}
\newacronym{ci}{CI}{Continuous Integration}
\newacronym{cd}{CD}{Continuous Deployment}
\newacronym{dtn}{DTN}{Data Transfer Node}

\newacronym{dns}{DNS}{Domain Name Service}
\newacronym{nrpe}{NRPE}{Nagios Remote Plugin Executor}
\newacronym{ldap}{LDAP}{Lightweight Directory Access Protocol}
\newacronym{lan}{LAN}{Local Area Network}
\newacronym{vlan}{VLAN}{Virtual LAN}

\newacronym{ipmi}{IPMI}{Intelligent Platform Management Interface}
\newacronym{tor}{ToR}{Top-of-the-Rack}
\newacronym{lmn}{LMN}{Large Memory Node}
\newacronym{bgp}{BGP}{Border Gateway Protocol}
\newacronym{dhcp}{DHCP}{Dynamic Host Configuration Protocol}
\newacronym{vrf}{VRF}{Virtual Routing and Forwarding}
\newacronym{vpn}{VPN}{Virtual Private Network}
\newacronym{rma}{RMA}{Return Merchandise Authorization}
\newacronym{hpc}{HPC}{High Performance Compute}

\newacronym{nu}{NU}{Northeastern University}
\newacronym{asic}{ASIC}{Application-specific Integrated Circuit}
\newacronym{rdma}{RDMA}{Remote Direct Memory Access}
\newacronym{roce}{RoCE}{RDMA over Converged Ethernet}
\newacronym{ovs}{OVS}{Open vSwitch}
\newacronym{frr}{FRR}{Free Range Routing}
\newacronym{ups}{UPS}{Uninterruptible Power Supply}

\newacronym{ntia}{NTIA}{National Telecommunications and Information Administration}
\newacronym{pii}{PII}{Personal and Identifiable Information}
\newacronym{irb}{IRB}{Institutional Review Board}
\newacronym{doi}{DOI}{Digital Object Identifier}

\newacronym{sdo}{SDO}{Standard-Development Organization}
\newacronym{ose}{OSE}{Open Source Ecosystem}
\newacronym{osc}{OSC}{O-RAN Software Community}
\newacronym{dop}{DOP}{Director of Operations}
\newacronym{pm}{PM}{Program Manager}
\newacronym{excom}{EXCOM}{Executive Committee}
\newacronym{iiot}{IIoT}{Industrial \gls{iot}}
\newacronym{lf}{LF}{Linux Foundation}

\newacronym{wiot}{WIoT}{Institute for the Wireless Internet of Things}

\newacronym{nofo}{NOFO}{Notice of Funding Opportunity}

\newacronym{onr}{ONR}{Office of Naval Research}
\newacronym{afosr}{AFOSR}{Air Force Office of Scientific Research}
\newacronym{afrl}{AFRL}{Air Force Research Laboratory}
\newacronym{arl}{ARL}{Army Research Laboratory}

\newacronym{arc}{ARC}{Aerial Research Cloud}
\newacronym{cast}{CaST}{Channel Emulation Scenario Generator and Sounder Toolchain}

\newacronym{mno}{MNO}{Mobile Network Operator}
\newacronym{ct}{CT}{Continuous Testing}
\newacronym{oci}{OCI}{Open Container Initiative}

\newacronym{xai}{XAI}{Explainable AI}
\newacronym{esc}{ESC}{Environmental Sensing Capability}
\newacronym{sas}{SAS}{Spectrum Access System}
\tikzstyle{startstop} = [rectangle, rounded corners, minimum width=2cm, minimum height=0.5cm,text centered, draw=black]
\tikzstyle{io} = [trapezium, trapezium left angle=70, trapezium right angle=110, minimum width=3cm, minimum height=1cm, text centered, draw=black]
\tikzstyle{process} = [rectangle, minimum width=2cm, minimum height=0.5cm, text centered, draw=black, alignb=center]
\tikzstyle{decision} = [ellipse, minimum width=2cm, minimum height=1cm, text centered, draw=black]
\tikzstyle{arrow} = [thick,<->,>=stealth]
\tikzstyle{line} = [thick,>=stealth]
\tikzstyle{darrow} = [thick,<->,>=stealth,dashed]
\tikzstyle{sarrow} = [thick,->,>=stealth]
\tikzstyle{larrow} = [line width=0.1mm,dashdotted,->,>=stealth]
\tikzstyle{llarrow} = [line width=0.1mm,->,>=stealth]

\makeatletter
\def\grd@save@target#1{%
  \def\grd@target{#1}}
\def\grd@save@start#1{%
  \def\grd@start{#1}}
\tikzset{
  grid with coordinates/.style={
    to path={%
      \pgfextra{%
        \edef\grd@@target{(\tikztotarget)}%
        \tikz@scan@one@point\grd@save@target\grd@@target\relax
        \edef\grd@@start{(\tikztostart)}%
        \tikz@scan@one@point\grd@save@start\grd@@start\relax
        \draw[minor help lines] (\tikztostart) grid (\tikztotarget);
        \draw[major help lines] (\tikztostart) grid (\tikztotarget);
        \grd@start
        \pgfmathsetmacro{\grd@xa}{\the\pgf@x/1cm}
        \pgfmathsetmacro{\grd@ya}{\the\pgf@y/1cm}
        \grd@target
        \pgfmathsetmacro{\grd@xb}{\the\pgf@x/1cm}
        \pgfmathsetmacro{\grd@yb}{\the\pgf@y/1cm}
        \pgfmathsetmacro{\grd@xc}{\grd@xa + \pgfkeysvalueof{/tikz/grid with coordinates/major step x}}
        \pgfmathsetmacro{\grd@yc}{\grd@ya + \pgfkeysvalueof{/tikz/grid with coordinates/major step y}}
        \foreach \x in {\grd@xa,\grd@xc,...,\grd@xb}
        \node[anchor=north] at (\x,\grd@ya) {\pgfmathprintnumber{\x}};
        \foreach \y in {\grd@ya,\grd@yc,...,\grd@yb}
        \node[anchor=east] at (\grd@xa,\y) {\pgfmathprintnumber{\y}};
      }
    }
  },
  minor help lines/.style={
    help lines,
    gray,
    line cap =round,
    xstep=\pgfkeysvalueof{/tikz/grid with coordinates/minor step x},
    ystep=\pgfkeysvalueof{/tikz/grid with coordinates/minor step y}
  },
  major help lines/.style={
    help lines,
    line cap =round,
    line width=\pgfkeysvalueof{/tikz/grid with coordinates/major line width},
    xstep=\pgfkeysvalueof{/tikz/grid with coordinates/major step x},
    ystep=\pgfkeysvalueof{/tikz/grid with coordinates/major step y}
  },
  grid with coordinates/.cd,
  minor step x/.initial=.5,
  minor step y/.initial=.2,
  major step x/.initial=1,
  major step y/.initial=1,
  major line width/.initial=1pt,
}
\makeatother

\definecolor{desireRed}{RGB}{230,57,60}%
\definecolor{darkPurple}{RGB}{59,31,43}%
\definecolor{springGreen}{RGB}{37,223,145}%
\definecolor{queenBlue}{RGB}{69,123,157}%
\definecolor{spaceCadet}{RGB}{29,53,87}%

\usepackage{dblfloatfix}

\newcommand{\coloran}{ColO-RAN\xspace}
\newcommand{\openrangym}{OpenRAN Gym\xspace}
\newcommand{\scope}{SCOPE\xspace}

\newcommand{\oran}{O-RAN\xspace}

\newcommand{\nearrt}{Near-\gls{rt}\xspace}

\newcommand{\ran}{\gls{ran}\xspace}

\newcommand{\ai}{\gls{ai}\xspace}
\newcommand{\xai}{\gls{xai}\xspace}

\newcommand*{\cicd}{\gls{ci}/\gls{cd}\xspace}

\begin{document}

\ifojtemplate
\else
\title{Colosseum: The Open RAN Digital Twin}
\fi

\ifojtemplate
\else

\author{\IEEEauthorblockN{Michele Polese, Leonardo Bonati, Salvatore D’Oro, Pedram Johari, Davide Villa,\\Sakthivel Velumani, Rajeev Gangula, Maria Tsampazi, Clifton Paul Robinson,\\Gabriele Gemmi, Andrea Lacava, Stefano Maxenti, Hai Cheng, Tommaso Melodia}
\thanks{The authors are with the Institute for the Wireless Internet of Things, Northeastern University, Boston, MA, USA. E-mail: m.polese@northeastern.edu.}
\thanks{This article is based upon work partially supported by the O-RAN ALLIANCE, the U.S.\ National Science Foundation under grants CNS-1925601 and CNS-2112471, by the National Telecommunications and Information Administration (NTIA)'s Public Wireless Supply Chain Innovation Fund (PWSCIF) under Awards No. 25-60-IF011 and 25-60-IF054, and by OUSD(R\&E) through Army Research Laboratory Cooperative Agreement Number W911NF-19-2-0221. The views and conclusions contained in this document are those of the authors and should not be interpreted as representing the official policies, either expressed or implied, of the Army Research Laboratory or the U.S. Government. The U.S. Government is authorized to reproduce and distribute reprints for Government purposes notwithstanding any copyright notation herein.}
}

\fi

\flushbottom
\setlength{\parskip}{0ex plus0.1ex}

\glsunset{nr}
\glsunset{lte}
\glsunset{3gpp}
\glsunset{cbrs}
\glsunset{usrp}

\ifojtemplate
\else
  \maketitle
\fi

\begin{abstract}
Recent years have witnessed the Open \gls{ran} paradigm transforming the fundamental ways cellular systems are deployed, managed, and optimized.
This shift is led by concepts such as openness, softwarization, programmability, interoperability, and intelligence of the network, all of which had never been applied to the cellular ecosystem before.
The realization of the Open \gls{ran} vision into practical architectures, intelligent data-driven control loops, and efficient software implementations, however, is a multifaceted challenge, which requires (i)~datasets to train \gls{ai} and \gls{ml} models; (ii)~facilities to test models without disrupting production networks; (iii)~continuous and automated validation of the \gls{ran} software; and (iv)~significant testing and integration efforts.
This paper poses itself as a tutorial on how Colosseum---the world's largest wireless network emulator with hardware in the loop---can provide the research infrastructure and tools to fill the gap between the Open \gls{ran} vision, and the deployment and commercialization of open and programmable networks.
We describe how Colosseum implements an Open \gls{ran} digital twin through a high-fidelity \gls{rf} channel emulator and end-to-end softwarized O-RAN and 5G-compliant protocol stacks, thus allowing users to reproduce and experiment upon topologies representative of real-world cellular deployments.
Then, we detail the twinning infrastructure of Colosseum, as well as the automation pipelines for \gls{rf} and protocol stack twinning.
Finally, we showcase a broad range of Open \gls{ran} use cases implemented on Colosseum, including the real-time connection between the digital twin and real-world networks, and the development, prototyping, and testing of \gls{ai}/\gls{ml} solutions for Open \gls{ran}.
\end{abstract}

\begin{IEEEkeywords}
O-RAN, Open RAN, Wireless Network Emulation, 5G, 6G.
\end{IEEEkeywords}

\maketitle

\begin{figure*}[t]
  \centering
  \includegraphics[width=\textwidth]{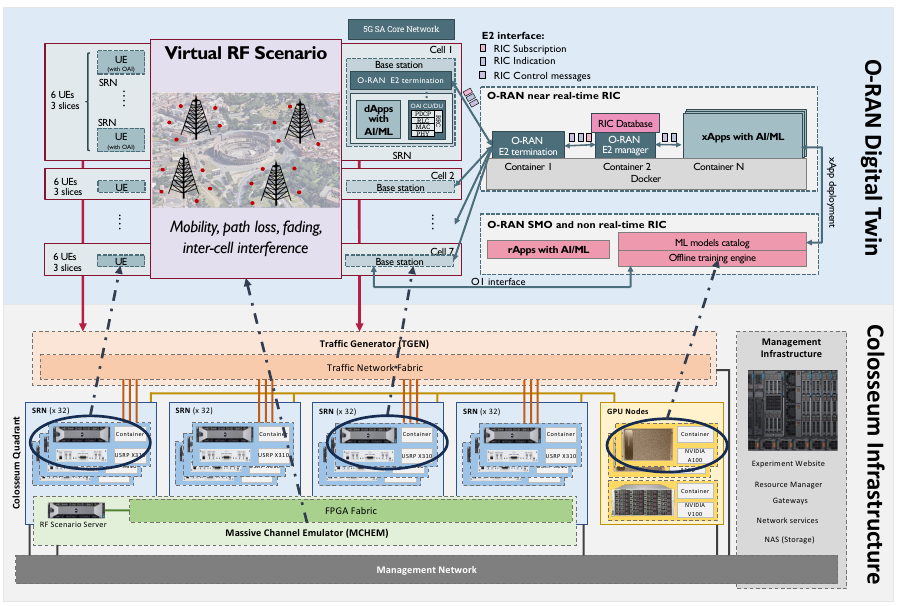}
  \caption{Open RAN twinning capabilities in Colosseum}
  \label{fig:colosseum-o-ran}
\end{figure*}

\glsresetall
\glsunset{nr}
\glsunset{lte}
\glsunset{3gpp}
\glsunset{cbrs}
\glsunset{usrp}

\ifojtemplate
\else
\section{Introduction}
\fi
\label{sec:introduction}

Wireless networks and systems are evolving into more programmable, open, and virtualized solutions. The transition to software-defined networking~\cite{mckeown2008openflow}, which has taken over the management operations of the wired networking industry in the last decade, is now moving to cellular networks.
While introducing software and programmability in radio systems is more challenging compared to switching and routing, it also opens opportunities to deploy and configure bespoke networks that can be dynamically tuned to optimally support a variety of different verticals.
The Open \gls{ran} paradigm, its embodiment in the technical specifications of the O-RAN ALLIANCE, and evolution in the 3GPP \gls{ran} and core network designs are converging toward a network architecture which is open, disaggregated, programmable, and intelligent~\cite{polese2022understanding,abdalla2022toward}.
Openness, embedded in interfaces that expose \gls{ran} telemetry and control, enables the programmability and optimization of \gls{ran} functions through intelligent closed-loop control driven from the so-called \glspl{ric}~\cite{oran-wg3-ricarch,balasubramanian2021ric}. Disaggregation, implemented by splitting \gls{ran} functionalities into self-contained units across user and control planes,
fosters innovation and introduces multi-vendor interoperability.
Finally, programmability, realized through open \glspl{api} of the software solutions, enables swift reconfiguration of the white-box components of the cellular networks to adapt to ever-changing network conditions and demand.
%
Overall, these ingredients have the potential to transform how we deploy, manage, and optimize cellular networks.

However, advancing the Open \gls{ran} vision toward a fully developed architecture with robust and reliable algorithmic components and seamless multi-vendor integration requires addressing several challenges at the architectural, algorithmic, and system-level design, which include the following.

\textbf{Need for Datasets.} To develop robust and scalable \gls{ai} and \gls{ml} solutions, which generalize well across a variety of real-world deployment scenarios, it is necessary to leverage rich datasets of \gls{ran} telemetry, data, and performance indicators~\cite{challita2020when}. While network operators are in a unique position to collect such datasets, it is often impractical or impossible to use them for research and development due to privacy and security concerns. Wireless testbeds represent a feasible path to overcome this limitation~\cite{villa2024dt,breen2020powder,raychaudhuri2020cosmos,marojevic2020advanced,zhang2021ara}. However, they are often limited to the \gls{rf} characteristics and topology of their deployment area. 

\textbf{End-to-end \gls{ai} and \gls{ml} Testing.} Once trained, \gls{ai}/\gls{ml}-based control solutions need to be validated and tested in controlled environments to avoid disruption in production networks. At the same time, the testing conditions need to be realistic to obtain meaningful results that consider, for instance, the user load, traffic patterns, and \gls{rf} characteristics of real-world deployments the models will be used to control.

\textbf{Continuous Software Validation.} While softwarization introduces flexibility and programmability of the stack, it also comes with concerns around software quality, reliability, security, and performance~\cite{hung2024security}. Therefore, integrating, validating, testing, and profiling software for wireless in a continuous fashion is key to the Open \gls{ran} vision. Additionally, this validation needs to consider various compute platforms and hardware acceleration solutions for physical layer processing.

\textbf{Automated Integration and Testing of Disaggregated Components.} Disaggregation comes with a more robust supply chain, but also a need for the validation of interoperability across vendors and devices. This is a labor-intensive and often manual process that calls for the development of automated techniques~\cite{dellOroRAN}.

In this paper, we describe how Colosseum---the world's largest wireless network emulator with hardware in the loop~\cite{bonati2021colosseum}---can be leveraged as a digital twin platform to address these challenges and to develop end-to-end, fully integrated, and reliable solutions for Open \gls{ran}, as shown in Figure~\ref{fig:colosseum-o-ran}.
Through its channel and traffic emulation capabilities, Colosseum can replicate countless real-world scenarios representative of real-world cellular deployments, and generate datasets that can be used to train robust \gls{ai}/\gls{ml} models robust to network changes~\cite{polese2021coloran}.
A combination of generic compute nodes, \glspl{sdr}, and the possibility of integrating \gls{cots} devices, allows for the digital replica of Open \gls{ran} and 5G-and-beyond protocol stacks~\cite{kaltenberger2020openairinterface,gomez2016srslte}, which we manage through automation and \cicd pipelines~\cite{bonati20235g}.
This also enables repeatable experiments, where different network configurations and protocol stacks can be tested against the same channel and traffic conditions, as well as a safe playground for testing of \gls{ai}/\gls{ml}solutions.

The remaining of this paper---which is meant as a tutorial that introduces Colosseum as an Open \gls{ran} digital twin---is organized as follows.
Section~\ref{sec:colosseum-intro} introduces Colosseum, its infrastructure, and capabilities, including recent extension that introduced \gls{gpu} servers for \gls{ai} and \gls{ml} training.
Section~\ref{sec:oai-o-ran} reviews the open-source protocol stacks that can be twinned in Colosseum to replicate 5G and Open RAN systems. These include \gls{oai}~\cite{kaltenberger2020openairinterface}, srsRAN~\cite{gomez2016srslte}, and the OpenRAN Gym~\cite{bonati2022openrangym-pawr} framework.
Section~\ref{sec:rf-twinning} reviews methods and techniques to replicate real-world \gls{rf} scenarios with high fidelity~\cite{villa2022cast,villa2024dt}, and discusses the planned evolution of the Colosseum channel emulator toward a dynamic \gls{rf} twinning.
Section~\ref{sec:automation} details the fully automated framework that we developed and that leverages \cicd techniques to continuously update and test the 5G and Open RAN protocol stacks in Colosseum. This section also presents a novel software broker to enable twinning of scenarios across the real and digital domains.
Section~\ref{sec:external} discusses how Colosseum experiments can extend to external platforms, including over-the-air-testbeds at Northeastern University~\cite{villa2024x5g,bertizzolo2020arena} and the \gls{pawr} platforms~\cite{breen2021powder,raychaudhuri2020cosmos}.
Section~\ref{sec:research} reviews Open \gls{ran} applications and use cases that have been demonstrated on Colosseum. These include intelligent network slicing~\cite{polese2021coloran,tsampazi2023comparative}, real-time spectrum sharing enabled by Open RAN interfaces and controllers~\cite{villa2023twinning,doro2022dapps}, explainable \gls{ai} for O-RAN~\cite{explora2023conext}, \gls{iab} optimization~\cite{moro2023toward}, and jamming and fingerprinting with O-RAN primitives.
Section~\ref{sec:otic} overviews the role of the Colosseum in the testing and integration of interoperable Open \gls{ran} systems.
Section~\ref{sec:soa} reviews the state of the art on digital twins and experimental frameworks for Open \gls{ran}.
Finally, Section~\ref{sec:conclusions} concludes our paper.
Overall, we believe that this tutorial paper can serve the research community as a comprehensive resource to approach, understand, and leverage the resources that Colosseum makes available in the context of Open \gls{ran} systems.

\ifojtemplate
\else
\section{Colosseum Wireless Network Emulator}
\fi
\label{sec:colosseum-intro}

The architecture of Colosseum is shown in the bottom part of Figure~\ref{fig:colosseum-o-ran}.
Its main components are: (i)~128~pairs of generic compute servers and \glspl{sdr}, named \glspl{srn}; (ii)~a channel emulation system; and (iii)~a state-of-the-art \gls{ai}/\gls{ml} infrastructure.

Each \gls{sdr} is paired with a Dell server connected to an NI/Ettus \gls{usrp} X310 \gls{sdr} through a dedicated $10$\:Gbps link.\footnote{We are in the process of updating our compute solutions within Colosseum, transitioning from Dell PowerEdge R730 servers to Dell PowerEdge R750.}
The users of Colosseum can reserve these \glspl{srn} and deploy either custom or pre-configured protocol stacks through \gls{lxc}.
Container images are stored in a dedicated storage infrastructure, which is part of a larger management infrastructure that includes services such as websites, \gls{vpn} for external connectivity, and networking functionalities.

The Colosseum channel emulator, namely \gls{mchem}, is in charge of reproducing digital representations of real-world \gls{rf} propagation scenarios, which represent the state of the \gls{rf} channel at given time instants.
Specifically, each pair of transceivers, i.e., those of the \glspl{sdr}, is modeled through a \gls{tdl} with up to four non-zero taps that represent the multipath components of the \gls{cir} for each millisecond of the captured scenario.
At its core, \gls{mchem} includes four quadrants with 64~\glspl{fpga} and 128~\glspl{sdr} that up-/down-convert the signal between \gls{rf} and baseband.
The \glspl{fpga} implement real-time convolutional operations to the digital baseband signal, modeling \gls{rf} taps of the wireless channel effects of the real-world environments.
These taps, which have a maximum propagation delay of $5.12\:\mathrm{\mu}h$s~\cite{chaudhari2018scalable}, are streamed to the \glspl{fpga} from a dedicated \gls{rf} scenario server.
We refer readers to~\cite{bonati2021colosseum} for a comprehensive overview of the Colosseum architecture.

The \gls{ai}/\gls{ml} infrastructure of Colosseum, shown in yellow at the bottom of Figure~\ref{fig:colosseum-o-ran}, consists of two NVIDIA DGX A100 stations with 8~\glspl{gpu} each, providing $10$\:petaFLOPS of compute power and one Supermicro Superserver 8049U-E1CR4T with 6 NVIDIA V100 \glspl{gpu} and $3$\:TB of RAM.
%
%
Compared to the virtualization system of the \glspl{srn}, the \gls{gpu} nodes leverage Docker containers.
Resources on these nodes are managed through HashiCorp frameworks~\cite{hashicorp}, such as Nomad (for workload orchestration), Consul (for service discovery, configuration, and connectivity), and Traefik (for reverse proxy and load balancing), deployed in a redundant manner for fault tolerance.

\ifojtemplate
\else
\section{End-to-End Open RAN Twinning in Colosseum}
\fi
\label{sec:oai-o-ran}

The emulation capabilities of Colosseum can be leveraged to deploy, study, and profile wireless protocol stacks in controlled and repeatable environments.
As an example, Figure~\ref{fig:colosseum-o-ran} showcases the deployment of an end-to-end, fully programmable Open \gls{ran} cellular system in Colosseum.
The top-left portion of the figure depicts the \gls{ran}, which consists of core network and cellular base stations---either monolithic or disaggregated in \glspl{cu} and \glspl{du}---deployed on an emulated \gls{rf} deployment of interest and serving mobile \glspl{ue}.
The top-right portion of the figure, instead, shows the \gls{smo} and \oran \glspl{ric} that interface with the \gls{ran} base stations through open interfaces and augment their operations through \gls{ai}/\gls{ml} applications deployed therein, namely xApps and rApps.
We refer the readers to~\cite{polese2022understanding} for a detailed overview of the O-RAN architecture and functionalities.

In Colosseum, these O-RAN-aligned solutions are provided to the users through ready-to-use \gls{lxc} images that implement the functionalities made available through the \openrangym framework, which will be described next.

\subsection{Open-source Cellular Protocol Stacks}
\label{sec:stacks}

Colosseum enables experimentation with different open-source softwarized protocol stacks, such as \gls{oai}~\cite{kaltenberger2020openairinterface} for 5G \glspl{ran} and srsRAN~\cite{gomez2016srslte} for 4G ones. These protocol stacks---which are provided as ready-to-use images~\cite{colosseumImages}---are the same that can be used for over-the-air experiments with minor variations in the configuration parameters, discussed next.

\textbf{OpenAirInterface.}~\gls{oai} is an open-source project that provides a reference implementation of 3GPP cellular protocol stacks, including that of 5G NR. The \gls{oai}-based NR system consists of 5G \gls{ran} and core network that run on Linux-based general purpose compute platforms and can control \glspl{sdr}, e.g., the NI/Ettus USRPs deployed in Colosseum. The \gls{oai} \gls{ran} applications support the instantiation of \glspl{gnb} and \glspl{ue}.
The \gls{gnb} can operate in monolithic or disaggregated manner that leverages the \gls{cu}/\gls{du} split option. Both modes are available in Colosseum.
With the disaggregated base station, \gls{cu} and \gls{du} exchange data with control and user plane via the F1-C and F1-U interfaces, respectively, through the F1AP protocol.
The \gls{cu} can further be disaggregated into \gls{cucp} and \gls{cuup}. The control information between control plane and user plane nodes follows the E1AP protocol. 

Users of Colosseum can deploy 5G NR cellular networks on Colosseum through ready-to-use container images.
Compared to over-the-air environments, however, the extra components in Colosseum transmit and receive chains (e.g., those part of \gls{mchem}) require adjusting the timing advance during the establishment of the initial connection between \gls{gnb} and \glspl{ue}.
This can be done through the \texttt{ta} input argument when starting the base station application.


\textbf{srsRAN.}~For legacy 4G networks, Colosseum provides container images for the 4G version of srsRAN. Similarly to \gls{oai}, this open-source software allows users to instantiate experiments with softwarized \glspl{enb}, \glspl{ue}, and core network. Differently from 5G networks, which use \gls{tdd}, 4G systems are primarily based on \gls{fdd}, with the downlink and uplink in two separate frequency bands. In Colosseum, these need to be accommodated in two portions of the $80$\:MHz usable bandwidth through a custom configuration of the srsLTE \gls{enb}, which is already pre-configured in the \gls{lxc} images available to Colosseum users.


\subsection{OpenRAN Gym}
\label{subsection:openrangym}

%

\begin{figure}[t]
    \centering
    \includegraphics[width=\columnwidth]{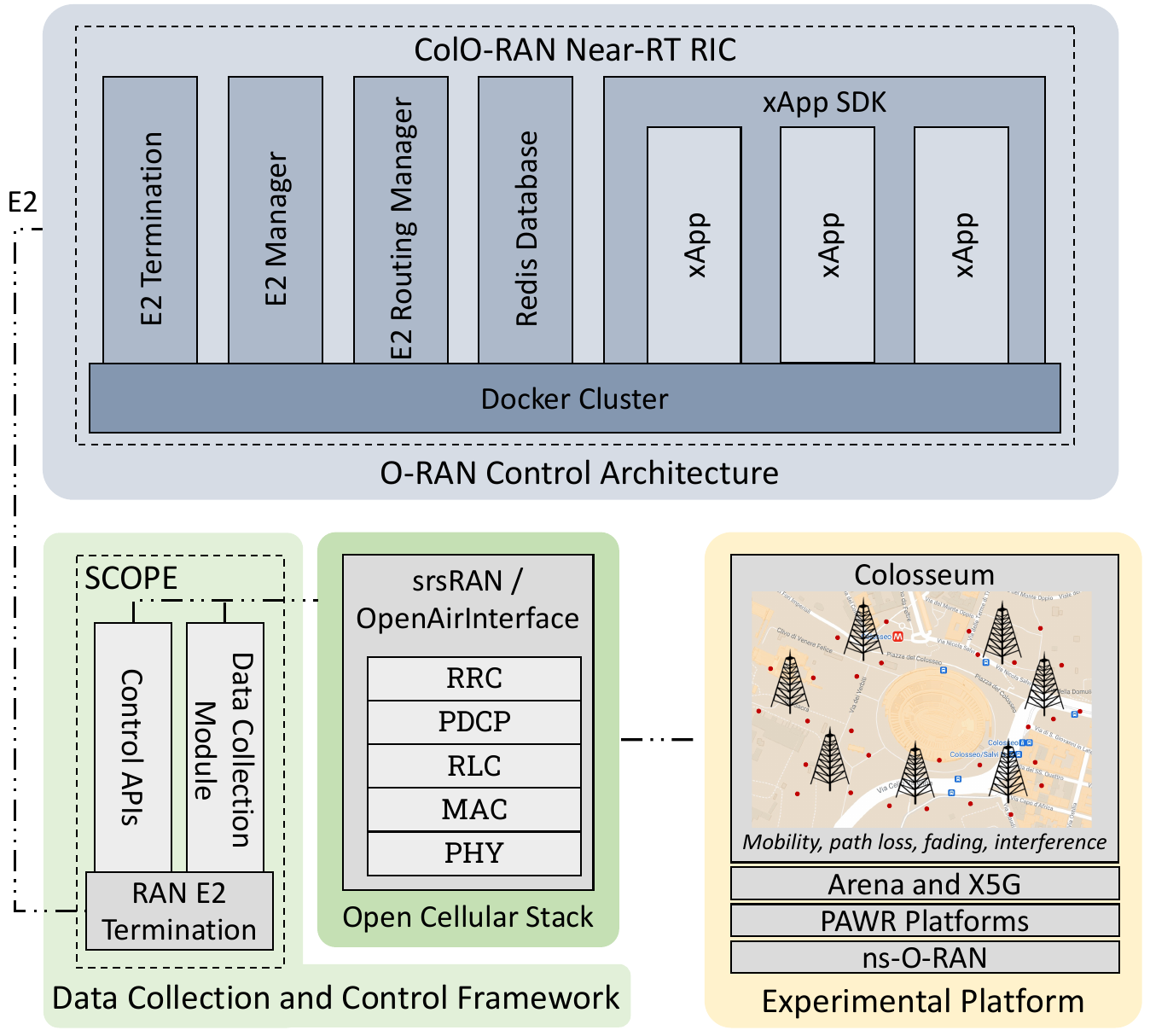}
    \caption{Intelligent Control Loops with Colosseum, based on the OpenRAN Gym framework~\cite{bonati2022openrangym-pawr}.}
    \label{fig:intelligent-loops}
\end{figure}


\openrangym is an open-source framework for O-RAN experimentation that lets users instantiate softwarized \glspl{ran} based on the \gls{oai} and srsRAN protocol stacks, and control them through xApps deployed on an O-RAN \nearrt \gls{ric}~\cite{bonati2022openrangym-pawr}.
Its components, shown in Figure~\ref{fig:intelligent-loops}, include (i)~\scope~\cite{bonati2021scope}, a wrapper for \gls{ran} software that simplifies the experiment workflow of users and enables data-collection at scale, and~(ii)~\coloran~\cite{polese2021coloran}, a simplified version of the O-RAN Software Community \nearrt \gls{ric} that provides an \gls{sdk} for xApp design, as well as pipelines for data collection and training of \gls{ai}/\gls{ml} models for \gls{ran} inference and control to be embedded into xApps.
Overall, \openrangym lets users instantiate a fully compliant \nearrt \gls{ric} using Docker containers, as well as xApps that can interact with the \gls{ran} (e.g., to receive \glspl{kpi} and enforce control policies) via the E2 interface.
Examples of developed xApps include \gls{drl} agents to optimize the performance of the \gls{ran} in real-time by controlling the scheduling and slicing configuration~\cite{polese2021coloran,tsampazi2023comparative,doro2022orchestran}. Tutorials on how to start using \openrangym on Colosseum are publicly available on the \openrangym website.\footnote{\openrangym website: \url{https://openrangym.com}.}

Finally, this framework has also been used to demonstrate the portability of experiments from Colosseum to external platforms, e.g., the Arena testbed~\cite{bertizzolo2020arena} and the platforms from the \gls{pawr} program~\cite{pawr} funded by the U.S.\ National Science Foundation, as discussed in Section~\ref{sec:colosseum-pawr-experiment}, and to interface with \glspl{ran} simulated through the ns-3 discrete-event network simulator using ns-O-RAN~\cite{10.1145/3592149.3592161}. This highlights the extensibility of the \openrangym framework and the flexibility of the digital twinning capabilities of Colosseum.


\ifojtemplate
\else
\section{Radio-Frequency Twinning}
\fi
\label{sec:rf-twinning}

The channel and traffic emulation and the cellular protocol stacks that can be deployed on Colosseum provide key components toward the creation of real-time, high-fidelity digital replicas of Open \gls{ran} systems with \gls{rf} hardware in the loop.
This section describes a set of toolchains that we designed and built to twin real-world \gls{rf} scenarios in Colosseum.
Section~\ref{sec:cast} describes our \gls{cast}, which enables \gls{rf} scenario twinning by pre-generating offline scenarios, while Section~\ref{sec:roadmap} discusses our roadmap toward real-time \gls{rf} twinning on Colosseum.

\subsection{RF Twinning with CaST}
\label{sec:cast}

\begin{figure}[htb]
    \centering
    \includegraphics[width=\columnwidth]{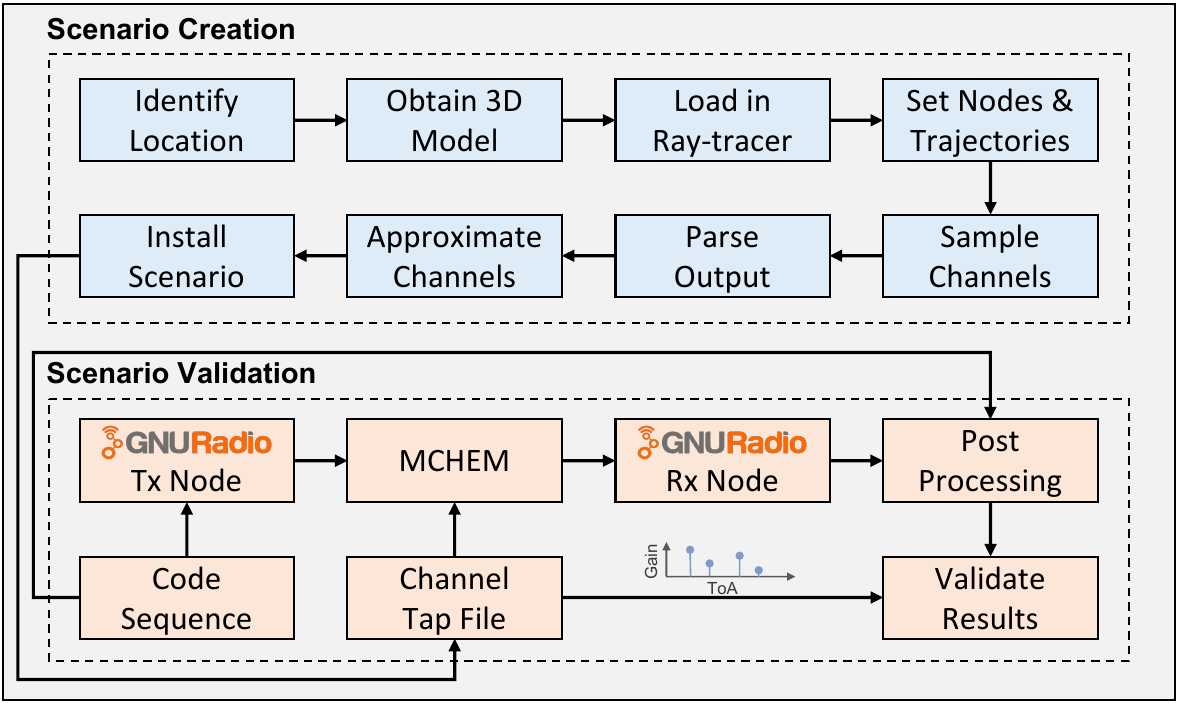}
    \caption{\gls{cast} scenario creation (top) and scenario validation (bottom) workflows.}
    \label{fig:cast-workflow}
\end{figure}

Because of the computational and time complexity of running ray-tracing operations to derive the taps of the wireless channels to emulate, \gls{rf} scenarios are currently generated offline, and then replayed in real-time by the scenario server.
To do so, we leverage the 
\gls{rf} twinning procedure of the \gls{cast} framework~\cite{villa2022cast}.
This toolchain, which is publicly available for the research community,\footnote{CaST is available at \url{https://github.com/wineslab/cast}.} enables users to characterize real-world \gls{rf} environments and to turn them into digital twin representations to be used in channel emulators such as Colosseum.
\gls{cast} workflow and procedures, depicted in Figure~\ref{fig:cast-workflow}, comprise two main components: (i)~a streamlined framework to create realistic mobile wireless scenarios (top part of the figure); and (ii)~an \gls{sdr}-based channel sounder for characterizing emulated \gls{rf} channels (bottom part). The combination of these two components allows for the generation of digital twin \gls{rf} scenarios, and for their validation, ensuring that the digital replicas of the wireless environments align with their real-world counterpart. 

\ifojtemplate
\else
\fi

The main steps of \gls{cast} scenario creation workflow are as follows:
(i)~identify the wireless environment to capture, i.e., the physical location to twin, which can vary in size and type, e.g., indoor, outdoor, urban, rural~\cite{villa2024dt, villa2023twinning};
(ii)~obtain the 3D model of the environment by leveraging online databases such as \gls{osm2} or modeling software such as SketchUp;
(iii)~load the model in a ray-tracer, e.g., MATLAB ray-tracer, \gls{wi}, Opal, or Sionna;
(iv)~set the location and trajectories of nodes;
(v)~sample the channel with the ray-tracer, capturing effects such as mobility of the nodes;
(vi)~parse the ray-tracer output to extract the channels for each pair of nodes for each millisecond of the emulated scenario;
(vii)~approximate the resulting channel information in a format suitable for the Colosseum channel emulator, i.e., 512 channel taps with at most 4 non-zero, spaced at $10$\:ns intervals and a maximum delay spread of $5.12\:\mathrm{\mu s}$;
(viii)~install the scenario in Colosseum, converting it into an \gls{fpga}-based format.
This scenario creation toolchain is modular, and it allows users to provide their inputs at any of the above-mentioned steps.

After a scenario is installed in Colosseum, it undergoes validation through the second part of the \gls{cast} framework to ensure that it closely follows the expected real-world behavior.
This involves a channel sounder implemented through the GNU Radio software~\cite{gnuradio}, and consists of the following steps:
(i)~transmission of a known code sequence used as a reference signal via the GNU Radio transmitter;
(ii)~reception via the GNU Radio receiver of the signal processed by \gls{mchem}, on top of which the channel characteristics were applied;
(iii)~post-processing of the received data with known code sequence with added channel effects to extract \gls{cir} and \gls{pl} of the waveform;
(iv)~validation of the results with the original modeled channel taps.
As demonstrated in~\cite{villa2022cast}, the \gls{cast} sounder is able to achieve an accuracy of $20$\:ns for \gls{cir} delays and $0.5$\:dB for tap gains. More details on the sounding sequences and processing algorithms can be found in~\cite{villa2022cast}.

Finally, since Colosseum \glspl{sdr} are equipped with two transmit-receive chains, Colosseum and \gls{mchem} support 2-by-2 \gls{mimo} operations.\footnote{Larger \gls{mimo} channels are possible, in principle, by connecting multiple \glspl{sdr} to the same compute server.}
Thus, \gls{cast} can be used to generate simplified \gls{mimo} channels leveraging MATLAB or Sionna's NR \gls{3gpp} channel models.
A comprehensive list of created and validated scenarios available to Colosseum users can be found in~\cite{colosseumscenariolist}.

\subsection{Roadmap Toward Real-Time Twinning}
\label{sec:roadmap}

At the time of this writing, Colosseum allows users to replay high-fidelity \gls{rf} scenarios that have been previously modeled offline and installed on the system.
While this process is extremely flexible and allows modeling \gls{rf} scenarios with different kind of tools (e.g., ray-tracers, mathematical models, etc.), it is not suitable for those cases in which the channel between nodes need to dynamically change in real-time.
This is the case, for instance, in which the digital twin needs to adapt to the state of the physical environment.

To overcome this limitation, we are updating the infrastructure of Colosseum to enable real-time \gls{rf} twinning.
Our phased approach combines an update in the \gls{rf} scenario server with a more flexible \gls{mchem} implementation, to stream channel taps generated in real-time, with real-time ray-tracing solutions.
In the following, we discuss possible approaches toward this, as well as some challenges.

\ifojtemplate
\else
\begin{figure*}
    \centering
    \includegraphics[width=\textwidth]{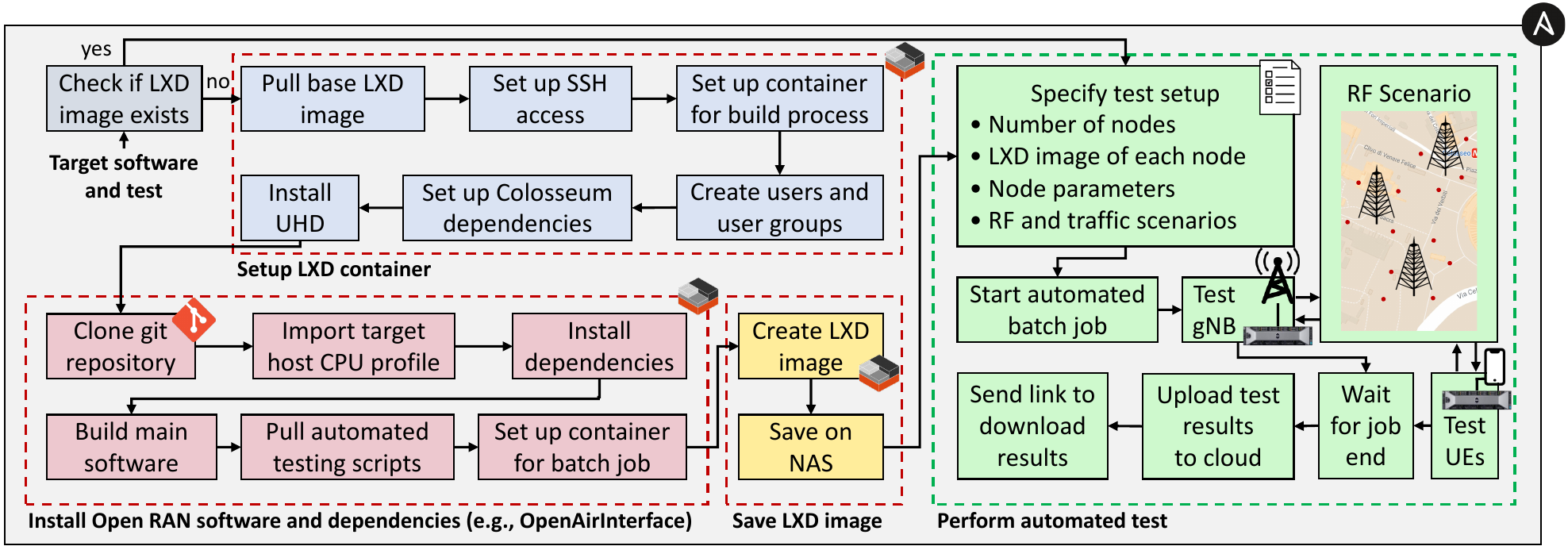}
    \caption{Ansible pipeline to perform an automated test on Colosseum.}
    \label{fig:ansible-pipeline-build-run}
\end{figure*}
\fi

\textbf{Real-time Channel Modeling.}
Using this approach, a detailed model of the \gls{ran} can be recreated in the digital twin using scenario and channel modeling techniques, e.g., 3D models and ray-tracing.
This model includes static and dynamic components.
The former is pre-configured in the system, and includes, for instance, the location of the base stations and buildings.
The latter is transmitted in real-time from the real-world \gls{ran} to reflect changes such as location of the \glspl{ue}, the transmission power.
%
%
Finally, once the digital component has been reconstructed, techniques such as ray-tracing or stochastic propagation models can be used to characterize the \gls{rf} propagation environment, then used to update \gls{mchem} in real-time.

\textbf{Real-time Channel Sensing.}
The previous approach can be extended with channel sensing techniques (e.g., the \gls{cast} sounding component) that continuously estimate channels across devices or test locations in the real-world environment.
This channel estimation can be continuously sent from the \gls{ran} to the digital twin, and then converted into an input that can be processed by \gls{mchem}.
Additionally, techniques could be used to derive an estimation of the channel taps from the \gls{csirs} or uplink \gls{srs} of the 5G NR physical layer~\cite{parkvall2017nr}.

\ifojtemplate
\else
\section{Automation for the Colosseum Digital Twin}
\fi
\label{sec:automation}

Besides the \gls{rf} and application twinning, a second component in the Colosseum digital twin platform allows for the automated replica of end-to-end, full-stack Open \gls{ran} systems across real-world testbeds and Colosseum. 
%
Colosseum implements \gls{ci}, \gls{cd}, and \gls{ct} pipelines for automating the twinning and testing of protocol stacks.
The \gls{ci} and \gls{cd} pipelines rely on Red Hat's Ansible automation framework, while the \gls{ct} process relies on Colosseum batch jobs.\footnote{Colosseum experiments can be run in interactive mode, where users are provided with a shell to the \gls{lxc} containers to run their code and experiments, or in batch mode, where the infrastructure takes care of executing the experiments whenever resources are available in the system~\cite{bonati2021colosseum}.}
A diagram of this workflow is shown in Figure~\ref{fig:ansible-pipeline-build-run}.

The \gls{ci}/\gls{cd} is formed of two logical steps: (i)~\textit{build a new image} (enclosed with dashed red lines), and (ii)~\textit{perform an automated test} with the built image (dashed green line).
Upon triggering the pipeline with a test to run, a check is made to ensure that an \gls{lxc} image of the software to test exists. In case of positive outcome from this check, the automated tests are performed; otherwise, a new \gls{lxc} image is built.
At a high level, the steps to build a new \gls{lxc} image involve three main tasks: (i)~\textit{setup a new \gls{lxc} container}, (ii)~\textit{install the Open \gls{ran} software and dependencies}, and (iii)~\textit{save the \gls{lxc} image}.
Setting up a new LXC container involves a series of steps that start with pulling a base LXC image, such as a base Ubuntu 22.04 image, from the repository with the public Colosseum LXC images. The setup process continues with the configuration of SSH access to allow users to log into the container once deployed on Colosseum. The container is then prepared for the build process by installing necessary compilers and tools, creating users and groups, and installing various dependencies and tools required by Colosseum. Additionally, the UHD software suite, essential for communication with the \glspl{sdr}, is installed.
Then, the task to install the Open \gls{ran} software and dependencies clones the repository of the software to test (e.g., the \gls{oai} software); imports the profile of the target node to build the software binary for (e.g., in case the compilation process is optimized for specific CPU instruction sets); installs the required dependencies, e.g., the ASN.1 definitions; builds the main software binaries; pulls some scripts to automated the testing process; and sets up the container for batch jobs execution on Colosseum.
Finally, the task to save the resulting \gls{lxc} image exports the container built so far as an \gls{lxc} image; and saves it on the Colosseum storage.

After building the \gls{lxc} image of the software to test, the step to perform automated tests through Colosseum batch jobs is run.
First, the test setup, which details the batch job parameters, is specified. These parameters include details on the number of nodes involved in the test, the \gls{lxc} image to load on each node (e.g., different nodes may act as \gls{gnb}, \gls{ue}, or core network, thus requiring different images), the \gls{ran} configuration (e.g., frequency and band to use), and the Colosseum \gls{rf} and traffic scenarios to run the test with. 
A new batch job is then triggered on Colosseum. This involves, for instance, deploying a \gls{gnb} providing service to cellular \glspl{ue}---both implemented via the \gls{oai} protocol stack---through heterogeneous Colosseum \gls{rf} scenarios.
After the batch job ends, results are uploaded to the cloud, and a link to download them is returned by the \gls{ci}/\gls{cd}/\gls{ct} pipeline so that users
can easily retrieve them.

\ifojtemplate
\else
\section{Connecting Real World and Digital Twin}
\fi
\label{sec:external}

The Colosseum digital twin can be integrated with external platforms and tools, either by connecting them directly (as described in Section~\ref{sec:colosseum-arena-integration} for the infrastructure part, and in Section~\ref{sec:mqtt} for the software enablers), or by transitioning experiments among platforms (Section~\ref{sec:colosseum-pawr-experiment}). In addition, it is possible to deploy and extend emulated experiments using discrete-event network simulators such as ns-3 and its O-RAN extensions~\cite{lacava2023programmable}.

\subsection{Integration with Arena and X5G}
\label{sec:colosseum-arena-integration}

The Arena testbed~\cite{bertizzolo2020arena} is an over-the-air indoor programmable testbed with 24 \glspl{sdr} connected to a grid of 64 antennas, enabling experimentation in the sub-$7.2$\:GHz frequency spectrum. It is deployed in the Boston campus of Northeastern University, and co-located with X5G, a 5G- and O-RAN-compliant testbed based on NVIDIA Aerial, which implements the NR physical layer on GPU, \gls{oai}, and commercial O-RAN \glspl{ru}~\cite{villa2024x5g}.
We leveraged the $10$\:Gbps shared metro link (round-trip time of $\sim$3~ms) between the Northeastern Boston and Burlington, MA campuses (where Colosseum is located) to set up a site-to-site \gls{vpn} between Colosseum and Arena, and to virtually merge the two networks through dedicated firewalls.
This integration will allow Colosseum users to execute experiments on the Arena testbed through unified web portal and \gls{lxc}, thus enabling swiftly transition of experiments between real and emulated environments.

\subsection{Cross-Twin Communication Broker for Scenario Twinning}
\label{sec:mqtt}

The Colosseum-Arena integration has also been used to prototype and test a software framework for the real-time twinning of a real-world scenario on Arena, and its digital counterpart on Colosseum.
We leveraged the MQTT protocol to implement the real-time communication between the two endpoints~\cite{standard2014mqtt}.
Because of its publish/subscribe mechanism, this protocol facilitates smooth data interchange among numerous devices, guaranteeing minimal bandwidth consumption, low power usage, and steadfast connectivity, which are key to our real-time twinning.
%

\begin{figure}[t]
    \centering
    \includegraphics[width=\columnwidth]{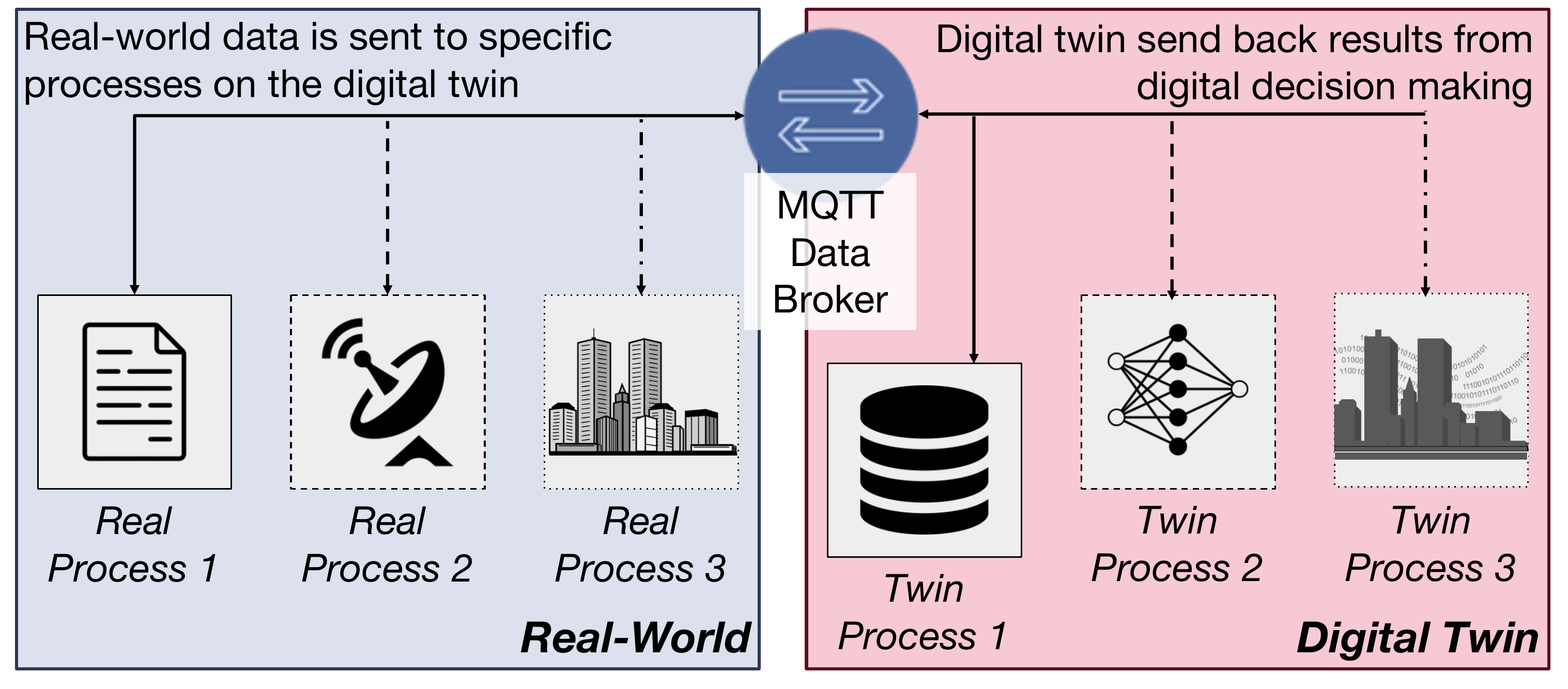}
    \caption{An overview depicting how the MQTT communication broker facilitates bidirectional data distribution to specific processes, enabling scalable implementation of digital twins.}
    \label{fig:mqtt-visual}
\end{figure}

The communication broker facilitates communication between devices and applications by managing the exchange of messages in a point-to-point manner. It allows for a single entity to control the flow of data in both directions, making management of the system streamlined, centralized, and responsive to changing data requirements.
We placed the broker in the digital twin environment, as shown in Figure~\ref{fig:mqtt-visual}, ensuring that resource constraints are not added to the real-world system, which acts as data distributor.
%

\begin{table}[h]
\setlength\belowcaptionskip{5pt}
    \centering
    \footnotesize
    \setlength{\tabcolsep}{2pt}
    \caption{Packet arrival times between the digital twin and the real-world counterpart in milliseconds~(ms). \textit{(Averaged latency of 100 samples)}}
    \label{tab:mqtt_dt}
    \begin{tabularx}{\columnwidth}{
        >{\raggedright\arraybackslash\hsize=0.8\hsize}X 
        >{\raggedright\arraybackslash\hsize=1\hsize}X
        >{\raggedright\arraybackslash\hsize=1\hsize}X }
        \toprule
        Packet Size & Real-to-Twin Latency & Twin-to-Real Latency \\
        \midrule
        1 Byte          & 15.318~ms     & 15.119~ms \\
        100 Bytes       & 15.547~ms     & 15.688~ms \\
        1 Kilobyte      & 15.742~ms     & 15.713~ms \\
        10 Kilobytes    & 22.261~ms     & 21.979~ms \\
        100 Kilobytes   & 31.101~ms     & 31.097~ms \\
        1 Megabyte      & 47.967~ms     & 48.003~ms \\
        \bottomrule
    \end{tabularx}
\end{table}

Table~\ref{tab:mqtt_dt} shows the communication latency between the two systems, showing how the packet size impacts the latency among the two systems.
As expected, smaller packets yield lower communication latency, e.g., in case of sensor reading or simple command exchanges.
Packet sized up to $100$\:byte show similar values for the latency. This could be the case, for instance, of short message transmissions.
Finally, larger packet sizes could support more substantial data transfers, e.g., small datasets spectrum information, while maintaining acceptable latency values.
%

\subsection{From Colosseum to the PAWR Platforms}
\label{sec:colosseum-pawr-experiment}

The capabilities of moving experiments between Colosseum and external testbeds have also been demonstrated in~\cite{bonati2022openrangym-pawr}, where an experimental campaign was first executed on Colosseum, and then transitioned to the Arena testbed~\cite{bertizzolo2020arena} and to the platforms of the \gls{pawr} program~\cite{pawr}.
Specifically, the \openrangym toolbox was leveraged on Colosseum to run a data collection campaign, train \gls{ml} xApps, and deploy them on a \nearrt \gls{ric} to perform data-driven control of a softwarized \gls{ran} implemented through the \scope framework and instantiated on an emulated urban \gls{rf} scenario~\cite{polese2021coloran, bonati2021scope}.
Then, the same xApps and \gls{lxc} containers used on Colosseum were transferred to the \gls{pawr} platforms and used to perform data-driven control of \glspl{ran} in both outdoor and indoor environments, achieving comparable trends in real-world and emulated environments.
This demonstrated the feasibility of transferring experiments between Colosseum and the external testbed.


\ifojtemplate
\else
\section{Use Cases of Colosseum as Open RAN Digital Twin}
\fi
\label{sec:research}


Colosseum has been extensively used to study open problems around the algorithmic, architectural, and system-level design of Open \gls{ran} systems, leveraging the realism in the \gls{rf} scenario emulation and the possibility of running state-of-the-art software-defined protocol stacks that twin real-world scenarios. In this section, we review recent results as examples of Open \gls{ran} studies that can be performed on Colosseum.

\subsection{Slicing and Scheduling Optimization}

%
\gls{ran} slicing allows operators to dynamically partition the available spectrum bandwidth to accommodate \glspl{ue} and traffic with heterogeneous \gls{qos}, performance, and \gls{sla} requirements.
A common example of \gls{ue} and type of service in 5G is that of \gls{embb}, \gls{mtc}, and \gls{urllc}.
Another control configuration that we considered is the selection of a different scheduling policies for each of the above-mentioned \gls{ran} slices~\cite{bonati2021intelligence, polese2021coloran}. 


To achieve flexibility in \gls{ran} control, \gls{ai}/\gls{ml} agents can be designed and embedded in xApps, then deployed on the \nearrt \gls{ric} and used to adapt the \gls{ran} to different network conditions and demand.
Specifically, \gls{drl} can be leveraged in the design of such control solutions for the Open~\gls{ran} as they can be trained a-priori and do not require prior knowledge of the run-time network dynamics~\cite{tsampazi2023comparative}.
Since these solutions require large datasets to be trained before they are deployed on a production \gls{ran} (to avoid disruption to its operations), the scale and controlled environment of digital twin offer fertile ground for their development.

\begin{figure}[t]
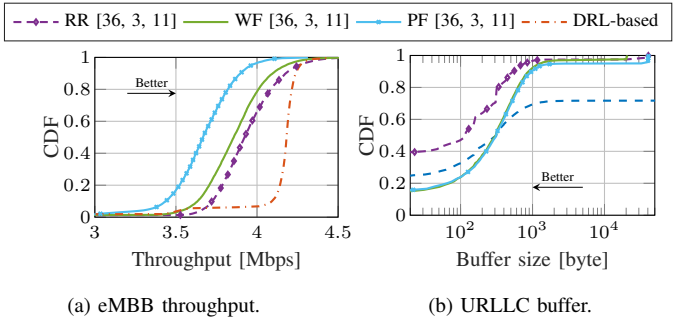

  \centering
  \begin{subfigure}[t]{0.48\columnwidth}
    \setlength\fwidth{.8\columnwidth}
    \setlength\fheight{.5\columnwidth}
    \input{figures/sched-slicing-model-comparison-th-slice-0.tex}
    \caption{eMBB throughput.}
    \label{fig:thr-embb}
  \end{subfigure}\hfill
  \begin{subfigure}[t]{0.48\columnwidth}
    \setlength\fwidth{.8\columnwidth}
    \setlength\fheight{.5\columnwidth}
    \input{figures/sched-slicing-model-comparison-buffer-slice-2.tex}
    \caption{\gls{urllc} buffer.}
    \label{fig:buf-urllc}
  \end{subfigure}
  \caption{Performance of a \gls{drl} agent controlling slicing and scheduling policies, from~\cite{polese2021coloran}. We compare the data-driven xApp and baselines without \gls{drl}-based adaptation. For the latter, the performance is based on the slicing configuration chosen with the highest probability by the best-performing \gls{drl} agent, and the three scheduler policies.}
  \label{fig:sched-slicing}
\end{figure}

Within this context, examples of Colosseum as the platform for the development of slicing and scheduling solutions for \oran systems are introduced in~\cite{polese2021coloran,bonati2022openrangym-pawr,tsampazi2023comparative,raftopoulos2024drl,moro2023toward,puligheddu2023semoran,bonati2021intelligence}.
Through the \openrangym framework (described in Section~\ref{subsection:openrangym}), Colosseum provides the tools and infrastructure for data collection, design, training, testing, and evaluation of \gls{ai}/\gls{ml} xApps before their deployment on over-the-air deployments.

As an example of this, in~\cite{polese2021coloran} we developed, trained, and tested \gls{drl}-based xApps for slicing and scheduling control of a softwarized \gls{ran} on the Colosseum digital twin.
The effectiveness of one of these xApps in controlling the scheduling (to be chosen among Round-robin (RR), Waterfilling (WF), and Proportionally Fair (PF)) and slicing (in terms of allocated \glspl{prb}) is shown in Figure~\ref{fig:sched-slicing}.
Specifically, Figure~\ref{fig:thr-embb} shows the throughput performance of an \gls{embb} slice, and Figure~\ref{fig:buf-urllc} the buffer occupancy of an \gls{urllc} slice, with our adaptive \gls{drl}-based xApp and static baselines.
We notice that the adapting the \gls{ran} configuration in near-real time allows the \gls{drl}-based xApp to outperform the baselines for the \gls{embb} slice, while matching the best static configuration for \gls{urllc} one.
The xApps trained in this way were then transitioned to the Arena testbed, where they adapted to over-the-air setup therein with only a few rounds of online learning, and to the \gls{pawr} platforms, where they exhibited similar trends to those achieved in Colosseum~\cite{bonati2022openrangym-pawr}.

%

Finally, in~\cite{tsampazi2023comparative}, we further expanded on this analysis by benchmarking $12$~xApps trained with different reward functions, action spaces, and hierarchical control.
These xApps demonstrated how different design choices might deliver higher performance while others might result in competitive behavior between different classes of traffic with similar objectives.




\subsection{Spectrum Sharing}

Colosseum has also been used to evaluate spectrum-sharing solutions for Open RAN.
These involve increased awareness of spectrum incumbents external to the \gls{ran}, as well as additional control actions (e.g., vacate a portion of the spectrum or null resource blocks).
The digital nature of the systems allows for a safe evaluation of sharing and coexistence schemes, without risks of harming incumbents, before the solutions are transitioned into the real world.

\begin{figure}[htb]
    \centering
    \includegraphics[width=\columnwidth]{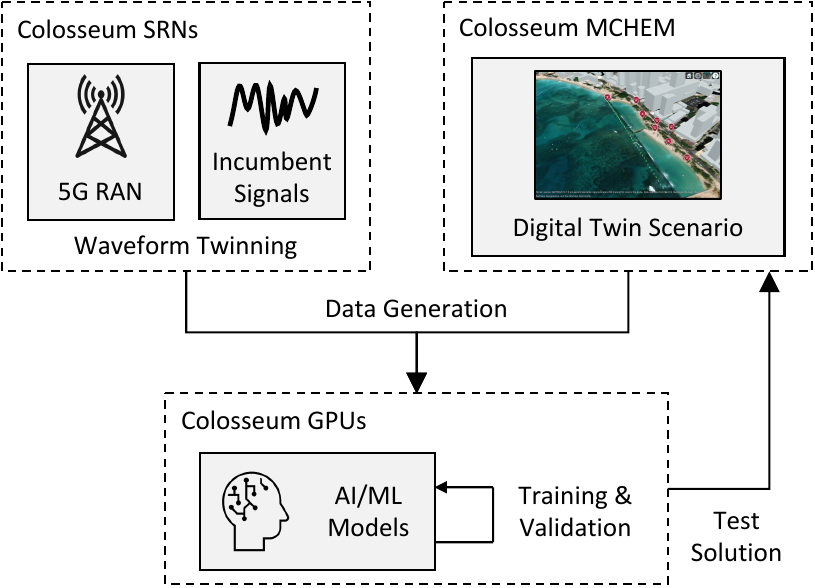}
    \caption{Colosseum workflow for data-driven spectrum-sharing solutions development and testing.}
    \label{fig:spectrum-sharing}
\end{figure}

Figure~\ref{fig:spectrum-sharing} shows an example of workflow that Colosseum enables for spectrum-sharing studies.
This includes steps such as waveform and scenario twinning, testing, and validation of the designed sharing solutions.

The papers~\cite{baldesi2022charm,villa2023twinning} leverage this workflow to study \gls{ai}-based sharing solutions, with the goal of improving
\gls{ran} performance over static schemes or model-based approaches.
%
Specifically, \cite{baldesi2022charm} leveraged Colosseum to collect a dataset with Wi-Fi and cellular users coexisting in the same frequency band, and to train a detector to identify spectrum usage patterns based on recurrent neural networks.
%
%
Similarly, \cite{villa2023twinning} demonstrated the use of Colosseum to carry out safe experiments with a radar incumbent operating in the \gls{cbrs} band~\cite{caromi2018detection}, and to vacate cellular operations from such band if incumbent are present.

\subsection{Explainable AI for Open RAN}


Although \ai will be a cornerstone technology for the Open \ran, one question that is still unresolved is how to guarantee that \ai solutions embedded into xApps and rApps are reliable enough to make autonomous decisions without the need for any human-in-the-loop.
Indeed, xApps and rApps can make decisions faster and more efficiently than humans thanks to the data-driven logic they embed. However, these decisions might result in outages and sub-optimal performance if \ai algorithms 
are not capable of adapting to changing conditions, and to compute network policies that are robust against noise and highly stochastic environments. 

In this context, \xai has emerged as a technology that offers a platform to understand \ai-based decision-making. Specifically, \xai makes it possible to explore relationships between outputs of \ai algorithms (e.g., control actions, predictions, forecasts) and inputs. Being able to study interactions between outputs and inputs of \ai solutions, \xai provides the necessary mechanisms to analyze \ai algorithms embedded in xApps and rApps and validate their decision-making. \xai can be helpful in many different ways as follows:
\begin{itemize}
    \item \textit{Performance parity:}~the most important aspect to bring \ai to fruition is performance parity against legacy \ran systems. Specifically, it is necessary to demonstrate that xApps and rApps can not only achieve performance parity, but also go beyond. \xai offers tools to identify conditions where \ai algorithms are under-performing and compare \ai outputs against baselines and expert knowledge.

    \item \textit{Robustness against noise and attacks:}~thanks to the ability to build explainable relationships between inputs and outputs, \xai makes it possible to determine the robustness of \ai solutions against noise, stochastic environments, and attacks. For example, \xai can be used to identify vulnerable portions of \ai algorithms (e.g., a certain layer, an activation function) that are subject to input fluctuations which can be either generated by noisy environments or exploited by attackers aware of the vulnerability.

    \item \textit{Design improvement:}~upon detecting pitfalls and weaknesses of the \ai algorithms as discussed above, these insights can be used to target portions of the \ai architecture that are under-performing (or that are vulnerable to noise or attacks) and fine-tune them to make them more robust and reliable.
\end{itemize} 

Colosseum can be used to design, develop, and test \xai techniques for \oran applications. Specifically, Colosseum is unique in that it can facilitate \xai development as it: (i)~supports testing of xApps and rApps across a variety of topology, mobility, channel, and traffic conditions to analyze their behavior across a variety of scenarios; (ii)~offers a ready-to-use development framework to integrate \xai algorithms into xApps and rApps; and (iii)~provides a realistic Open \gls{ran} digital twin that enables safe \xai analysis of xApps and rApps prior to their actual deployment on the production network, so as to avoid potential performance degradation and/or outages.
For example, the EXPLORA framework, presented in~\cite{explora2023conext}, is an \xai tool for \gls{drl}-based xApps and rApps developed on Colosseum via \openrangym. EXPLORA generates high-level interpretable explanations that describe why certain actions have been selected by the \gls{drl} agent, as well as operational conditions (e.g., traffic conditions) that led to a certain choice. Moreover, the \xai insights generated by EXPLORA can be used to identify sub-optimal action-state pairs and prevent the enforcement of such actions by replacing them with other actions that deliver higher performance.

\subsection{\gls{iab} Optimization with Open RAN}

\gls{iab} is a flexible wireless backhaul technology for cellular systems that has proved effective in reducing the deployment costs for ultra-dense \glspl{ran}~\cite{madapatha2020integrated,saha2019millimeter}. It was introduced in the 5G NR standard specifications as part of \gls{3gpp} Release~16, and comes with a new kind of \gls{ran} entity, called \gls{iab}-node.
The \gls{iab} node includes a \gls{mt} for wireless backhaul communications with upstream nodes, as well as the lower part of the cellular protocol stack (i.e., the layers in the \gls{du} and \gls{ru}). 
\gls{iab} nodes are not connected through wired paths to the core network, instead, they interconnect to each other creating a multi-hop graph that leverages the same waveform, protocol stack, and potentially the same spectrum. This wireless graph terminates in base stations with wired connections to the core network, called \gls{iab} donors.

Colosseum has been used as a digital playground to prototype and test in-band \gls{iab} networks based on \gls{oai}.
While not fully standard compliant, the solution developed in~\cite{moro2023toward} allows for the deployment of \gls{iab} nodes and donors with minimal modifications to \gls{oai}.
This enables to study relevant problems and use cases in the \gls{iab} domain, such as backhaul and access traffic partitioning, latency in multi-hop \gls{iab}, and optimal routing and scheduling.
Since the wireless nature of the backhaul link in \gls{iab} increases the space and parameters for network control, the Open \gls{ran} architecture and the \glspl{ric} can be leveraged to orchestrate and optimize \gls{iab} deployments with a data-driven approach.



\subsection{Security in Open RAN}

Security is another key concern and research and development area within Open RAN systems. Virtualization, softwarization, and open interfaces extend the attack surface of cellular systems, calling for additional testing and scrutiny into network management procedures, protocols, and implementations~\cite{shen2022security, ramezanpour2022intelligent, 9604996, 10.1145/3495243.3558259,chiejina2024system}. At the same time, open interfaces allow for increased visibility into the dynamics and telemetry of these systems, opening the door to robust data-driven anomaly- and intrusion-detection schemes~\cite{wen5g,scalingi2024det}. Finally, as the 5G cellular technology becomes more prevalent, the risk of more complex jamming attacks escalates, threatening network availability, reliability, and security~\cite{lichtman20185g,houda2024federated}. This makes the case for continuing research on jamming to safeguard future communications from malicious interference. As a digital twin based on real-time realistic \gls{rf} emulation, Colosseum can be used as a sandbox to assess threats, test countermeasures, and validate protection against jammers without the risk of harming real-world over-the-air systems.

\textbf{Colosseum for Evaluating Open RAN Security.} The digital twinning capabilities of Colosseum have been used to evaluate the security of O-RAN interfaces and implementations in~\cite{groen2023implementing}. 
This assessment has been performed on the twinned 5G and Open \gls{ran} end-to-end protocol stacks deployed on the Colosseum emulation infrastructure in~\cite{groen2023cost}, focusing on the E2 interface and on numerically and experimentally evaluating the cost associated with supporting modern encryption techniques for the E2 SCTP stream.

Beside securing the interfaces, there is wide interest in evaluating how the openness of the O-RAN architecture makes it possible to design and implement more advanced anomaly- and intrusion-detection mechanisms.
In~\cite{groen2023tractor,belgiovine2024megatron}, the authors investigate traffic classification mechanisms for Open \gls{ran} systems, implemented on the \nearrt \gls{ric} for the analysis of specific \glspl{kpi} associated with end-to-end application patterns. The paper~\cite{scalingi2024det} takes a step further and combines these insights with deep learning on the \gls{ran} I/Q samples (through a dApp~\cite{doro2022dapps}) to detect anomalies related to, for example, cloned user secure identity modules.

\textbf{Colosseum for Evaluating Jamming Attacks.}
Understanding evolving jamming techniques is crucial to developing effective countermeasures and ensuring uninterrupted connectivity for critical services relying on 5G networks. However, often times laws and regulations prohibit causing intentional interference to authorized radio communication systems, which can significantly hinder large-scale jamming research. This opens the door for digital twins to offer improved jamming research due to the accurate, real-world emulation.

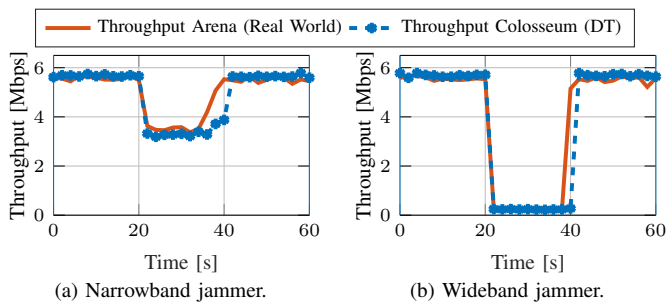
\begin{figure}[!t]
    \centering
   \begin{subfigure}[t]{.48\columnwidth}
    \setlength\fwidth{.8\columnwidth}
    \setlength\fheight{.5\columnwidth}
    \setlength\abovecaptionskip{-.01cm}
%
%
\definecolor{mycolor1}{rgb}{0.00000,0.44700,0.74100}%
\definecolor{mycolor2}{rgb}{0.85000,0.32500,0.09800}%
\begin{tikzpicture}
\pgfplotsset{every tick label/.append style={font=\scriptsize}}

\begin{axis}[%
width=\fwidth,
height=\fheight,
at={(0\fwidth,0\fheight)},
scale only axis,
xmin=0,
xmax=60,
xlabel style={font=\footnotesize\color{white!15!black}},
xlabel={Time [s]},
separate axis lines,
every outer y axis line/.append style={mycolor1},
every y tick label/.append style={font=\scriptsize\color{black}},
every y tick/.append style={black},
ymin=0,
ymax=6.5,
ylabel style={font=\footnotesize\color{black}},
ylabel shift=-5pt,
ylabel={Throughput [Mbps]},
axis background/.style={fill=white},
xmajorgrids,
ymajorgrids,
legend style={legend cell align=left, align=left, draw=white!15!black, font=\scriptsize, anchor=south, at={(1.1,1.05)}},
legend columns=2
]
\addplot [color=mycolor2, line width=1.5pt]
  table[row sep=crcr]{%
0	5.582\\
2	5.5576\\
4	5.445\\
6	5.6808\\
8	5.6252\\
10	5.5942\\
12	5.5172\\
14	5.514\\
16	5.5438\\
18	5.6444\\
20	5.508\\
22	3.6266\\
24	3.4714\\
26	3.4562\\
28	3.5668\\
30	3.5806\\
32	3.372\\
34	3.5114\\
36	4.1932\\
38	5.0702\\
40	5.532\\
42	5.4874\\
44	5.4264\\
46	5.6514\\
48	5.381\\
50	5.509\\
52	5.6262\\
54	5.5938\\
56	5.3434\\
58	5.523\\
60	5.4502\\
};
\addlegendentry{Throughput Arena (Real World)}

\addplot [color=mycolor1, dashed, line width=1.5pt, mark size=2.0pt, mark=asterisk, mark options={solid, mycolor1}]
  table[row sep=crcr]{%
0	5.616\\
2	5.6934\\
4	5.6866\\
6	5.6434\\
8	5.742\\
10	5.658\\
12	5.73\\
14	5.6514\\
16	5.642\\
18	5.694\\
20	5.6528\\
22	3.3224\\
24	3.1902\\
26	3.2698\\
28	3.2792\\
30	3.3094\\
32	3.218\\
34	3.3928\\
36	3.284\\
38	3.7134\\
40	3.8768\\
42	5.644\\
44	5.6314\\
46	5.6144\\
48	5.6694\\
50	5.6304\\
52	5.6578\\
54	5.627\\
56	5.6412\\
58	5.781\\
60	5.5922\\
};
\addlegendentry{Throughput Colosseum (DT)}

\end{axis}

\end{tikzpicture}%
    \caption{Narrowband jammer.}
    \label{fig:narrowband}
  \end{subfigure}\hfill
  \begin{subfigure}[t]{.48\columnwidth}
    \setlength\fwidth{.8\columnwidth}
    \setlength\fheight{.5\columnwidth}
    \setlength\abovecaptionskip{-.01cm}
%
%
\definecolor{mycolor1}{rgb}{0.00000,0.44700,0.74100}%
\definecolor{mycolor2}{rgb}{0.85000,0.32500,0.09800}%
\begin{tikzpicture}
\pgfplotsset{every tick label/.append style={font=\scriptsize}}

\begin{axis}[%
width=\fwidth,
height=\fheight,
at={(0\fwidth,0\fheight)},
scale only axis,
xmin=0,
xmax=60,
xlabel style={font=\footnotesize\color{white!15!black}},
xlabel={Time [s]},
separate axis lines,
every outer y axis line/.append style={mycolor1},
every y tick label/.append style={font=\scriptsize\color{black}},
every y tick/.append style={black},
ymin=0,
ymax=6.5,
ylabel style={font=\footnotesize\color{black}},
ylabel shift=-5pt,
ylabel={Throughput [Mbps]},
axis background/.style={fill=white},
xmajorgrids,
ymajorgrids
]
\addplot [color=mycolor2, line width=1.5pt]
  table[row sep=crcr]{%
0	5.5782\\
2	5.6862\\
4	5.6576\\
6	5.6154\\
8	5.4638\\
10	5.5562\\
12	5.5192\\
14	5.4966\\
16	5.5268\\
18	5.5644\\
20	5.5312\\
22	0.2344\\
24	0.2794\\
26	0.25\\
28	0.2778\\
30	0.2396\\
32	0.29\\
34	0.284\\
36	0.2776\\
38	0.2504\\
40	5.1496\\
42	5.5328\\
44	5.4626\\
46	5.6048\\
48	5.4154\\
50	5.4742\\
52	5.65\\
54	5.5574\\
56	5.6174\\
58	5.2154\\
60	5.581\\
};

\addplot [color=mycolor1, dashed, line width=1.5pt, mark size=2.0pt, mark=asterisk, mark options={solid, mycolor1}]
  table[row sep=crcr]{%
0	5.78\\
2	5.5878\\
4	5.782\\
6	5.706\\
8	5.686\\
10	5.632\\
12	5.6334\\
14	5.6946\\
16	5.664\\
18	5.71\\
20	5.712\\
22	0.248\\
24	0.235\\
26	0.2476\\
28	0.2302\\
30	0.2436\\
32	0.2266\\
34	0.2176\\
36	0.233\\
38	0.2366\\
40	0.2746\\
42	5.78\\
44	5.696\\
46	5.668\\
48	5.652\\
50	5.738\\
52	5.7454\\
54	5.654\\
56	5.738\\
58	5.668\\
60	5.6356\\
};

\end{axis}



\end{tikzpicture}%
    \caption{Wideband jammer.}
    \label{fig:wideband}
  \end{subfigure}
    \caption{Throughput results from jamming experiments on the Arena and Colosseum testbeds, showcasing narrowband and wideband cases, adapted from~\cite{villa2024dt}.}
    \label{fig:dt-jam-comp}
\end{figure}

Large-scale jamming experimentation offers a unique look at how jamming attacks impact the spectrum. These signals have specific attack goals, but there is also incidental interference that can occur in wireless communications. By emulating real-world environments through digital twins, it is possible to show both the intentional and accidental consequences of jamming attacks and gain crucial insight on them.
For example, in \cite{cprobinson_esword}, Colosseum has been used to perform large-scale controlled jamming experiments with hardware in the loop and a commercial jamming system. 
The experimental findings reveal that the digital twin achieves an accuracy of up to 98\% in replicating throughput, \gls{sinr}, and link status patterns when compared to real-world jamming experiments.
%
%
Additionally, Figure~\ref{fig:dt-jam-comp} compares results for two types of jamming attacks, narrowband and wideband, against an OFDM system.
The narrowband signal occupies $\sim156$\:kHz of the $20$\:MHz spectrum used for communications, the wideband signal occupies $10$\:MHz.
By looking at the figure, we notice that, in the real-world experiments, narrowband jamming causes throughput decreases of around 37-43\%, while wideband jamming of around 94-96\%.
Similar trends are observed in digital twin experiments. 

\ifojtemplate
\else
\section{Colosseum for O-RAN Testing and Integration}
\fi
\label{sec:otic}

The Colosseum Open \gls{ran} digital twin is also at the center of the recently established \gls{otic} in the Northeastern University Open6G center. An \gls{otic} is a center officially recognized by the O-RAN ALLIANCE for the testing and integration of Open \gls{ran} components from multiple vendors. Currently, the main testing and integration activities carried out by \glspl{otic} worldwide relate to achieving interoperability in the disaggregated Open \gls{ran} architecture. For instance, \glspl{otic} can assess the conformance of an O-RAN interface or piece of equipment with respect to the O-RAN specifications, or certify the interoperability of pairs of components and end-to-end systems.

While the Colosseum data center is currently equipped with solutions to perform such testing, we also consider its twinning capabilities---together with the controlled and repeatable experimental environment and its \gls{ai}/\gls{ml} infrastructure---as a key component in driving the testing and, eventually, certification of \gls{ai} solutions for the Open \gls{ran}. As of today, there are indeed discussions around interoperability certification and conformance of interfaces associated with the \glspl{ric}, but testing and evaluating \gls{ai} is an open challenge that requires further analysis and development.

\gls{ai} solutions may indeed be non-deterministic, and exhibit different behaviors when developed in different scenarios and under dynamic conditions and network load. This makes defining requirements for the certification of \gls{ai} solution a non-trivial problem. A possible strategy is leveraging benchmarking scenarios, as typically done with datasets in the computer vision community. In this sense, Colosseum, with its \openrangym framework, can implement and provide such reference scenarios to the community, leveraging a variety of wireless deployment scenarios, open-source protocol stacks, and either open-source or commercial \gls{ric} platforms. Testing can involve minimum accuracy or efficacy scores to achieve in the reference scenarios using xApps or rApps, adopting explainable \gls{ai} frameworks, and providing confidence levels around \gls{ai} agents misbehaving under unknown conditions. 

Finally, Colosseum is evolving to accommodate custom and \gls{cots} radios connected to the digital twin \gls{rf} emulation infrastructure. This makes it possible to onboard \gls{rf} equipment that allows for evaluations beyond the capabilities of \gls{usrp} \glspl{sdr}, e.g., O-RAN-compliant \glspl{ru} or commercial radars. Examples in the literature include studies with programmable Wi-Fi boards in~\cite{baldesi2022charm}, and jamming experiments in~\cite{villa2024dt,cprobinson_esword}.

\ifojtemplate
\else
\section{Related Work}
\fi
\label{sec:soa}

The development of digital twins and testbeds for cellular systems and, specifically, for Open \gls{ran} is a topic that has recently received significant attention, owing to the importance of addressing the challenges described in Section~\ref{sec:introduction}. The O-RAN ALLIANCE itself has dedicated a study area to digital twins for Open \gls{ran} within its next Generation Research Group (nGRG).

Digital twin frameworks for cellular networks are discussed in~\cite{jagannath2022digital,masaracchia2023digital,corici2023digital,nguyen2021digital,lin20236g,alkhateeb2023real,MCMANUS2023110000,mirzaei2023network}, with an analysis of the challenges and opportunities of using digital twins for Open \gls{ran} and beyond-5G design. 
Paper~\cite{vila2023design} proposes a network digital twin framework for 5G and develops a use case on \gls{ran} control with reinforcement learning, with a simulation-based twin that replicates a real-world scenario in Barcelona, Spain. The authors of~\cite{li2022rlops} discuss the opportunity of integrating digital twins in automated operations for the development of reinforcement learning algorithms for Open \gls{ran}. Reference~\cite{ndikumana2023digital} performs a simulation study and frames it in the context of digital twins to optimize fixed wireless access use cases for cellular networks. The studies in~\cite{deng2021digital,ren2023end} leverage a digital twin to create a virtual environment for the testing of network updates driven by \gls{ai}/\gls{ml} components. These papers, however, are primarily based on simulation-driven twins, which fails to capture the complexity of real-world \gls{rf} equipment. In addition, there is a limited evaluation of the digital twin in the real world, e.g., through experiments in real networks or testbeds. The Colosseum digital twin described in this paper is based on real-time channel emulation and end-to-end Open \gls{ran} protocol stacks that can be either automatically deployed and tested in the real world or in the digital twin as-is, thus achieving a high level of fidelity as testified by the experiments in~\cite{villa2024dt,bonati2022openrangym-pawr}.

Implementations of digital twins or components of digital twins for networked systems are presented in~\cite{moorthy2022middleware,HU2023342,baranda2021demo}, focusing on middleware for digital twins for aerial networks and on optimizing network service for the replica of a real-world setup (e.g., for Industry 4.0) into a digital twin. These papers, however, do not focus on end-to-end Open \gls{ran} systems and on the development of intelligent control solutions. Testbeds and frameworks for the evaluation of Open \gls{ran} systems are discussed in~\cite{bahl2023accelerating,villa2024x5g,upadhyaya2023open,upadhyaya2022prototyping,koumaras20185genesis}, and Open \gls{ran} evaluation is also supported in the \gls{pawr} platforms~\cite{breen2020powder,raychaudhuri2020cosmos,marojevic2020advanced,zhang2021ara}. These systems can be (and have been) used in combination with the twinning infrastructure of Colosseum to replicate Open \gls{ran} studies across twinned and real-world domains. 

\ifojtemplate
\else
\section{Conclusions and Future Work}
\fi
\label{sec:conclusions}

This paper reviewed how Colosseum, the world's largest wireless network emulator with hardware in the loop, can be leveraged as a digital twin for end-to-end Open \gls{ran} systems. We introduced the Colosseum architecture and focused on how it enables a high-fidelity digital replica of real-world \gls{rf} environments, as well as the deployment of real-world protocol stacks, both through automated pipelines. We then discussed how the digital twin integrates with real-world testbeds, and introduced several use cases where Colosseum digital twin capabilities have been used to advance algorithms, architecture, and testing of Open \gls{ran} systems.

We showed how Colosseum, through its emulation capabilities and support for softwarized stacks, allows for the development of effective solutions for the control and optimization of the \gls{ran}, from slicing and radio resource management to spectrum sharing and explainable \gls{ai} techniques. 

Colosseum is constantly evolving to further improve its support for the design and evaluation of wireless systems. As future work, beside a refresh of the hardware infrastructure, we plan to introduce more flexible channel emulation options and to tighten and streamline the connectivity across the digital and real-world testbeds discussed in this work.

\footnotesize
\balance
\bibliographystyle{IEEEtran}
\bibliography{biblio}

\end{document}